\documentclass[]{interact}

\usepackage{epstopdf}
\usepackage{subfigure}

\usepackage[numbers,sort&compress]{natbib}
\bibpunct[, ]{[}{]}{,}{n}{,}{,}

\theoremstyle{plain}

\theoremstyle{definition}

\theoremstyle{remark}

\begin{document}


\title{An Introduction to the {\em Planck} Mission}

\author{
\name{David L. Clements\textsuperscript{a}\thanks{CONTACT D.L. Clements. Email: d.clements@imperial.ac.uk}}
\affil{\textsuperscript{a}Physics Department, Imperial College, Prince Consort Road, London, SW7 2AZ, UK}
}

\maketitle

\begin{abstract}

The Cosmic Microwave Background (CMB) is the oldest light in the universe. It is seen today as black body radiation at a near-uniform temperature of 2.73K covering the entire sky. This radiation field is not perfectly uniform, but includes within it temperature anisotropies of order $\Delta T/T \sim10^{-5}$. Physical processes in the early universe have left their fingerprints in these CMB anisotropies, which later grew to become the galaxies and large scale structure we see today. CMB anisotropy observations are thus a key tool for cosmology. The {\em Planck} Mission was the European Space Agency's (ESA) probe of the CMB. Its unique design allowed CMB anisotropies to be measured to greater precision over a wider range of scales than ever before. This article provides an introduction to the {\em Planck} Mission, including its goals and motivation, its instrumentation and technology, the physics of the CMB, how the contaminating astrophysical foregrounds were overcome, and the key cosmological results that this mission has so far produced.

\end{abstract}


\section{Introduction}

The Cosmic Microwave Background (CMB) was discovered by Arno Penzias and Robert Wilson using an absolute radiometer working at a wavelength of 7 cm \cite{pw65}. It was one of the great serendipitous discoveries of science since the radiometer was intended as a low-noise ground station for the Echo satellites and Penzias and Wilson were looking for a source of excess noise rather than trying to make a fundamental observation in cosmology. Nevertheless, like the famous detective, once they had eliminated all other sources for this excess noise, including the `white dielectric substance' left by visiting pigeons, the remaining option, that they had detected a uniform all-sky source of radiation, however unlikely, had to be the truth. Their discovery came at a turbulent time for cosmology, with  the proponents of a static, Steady State, universe and an evolving universe engaged in a hotly contested debate. So contentious was the issue that one of the chief Steady State advocates, Fred Hoyle, had derided the evolving universe model as the Big Bang theory. However, with Penzias and Wilson's discovery, rapidly interpreted as being the glow left behind by a hot, dense phase of the early universe \cite{d65}, it became clear that the Big Bang theory was correct.

From the perspective of present-day cosmology the Big Bang model is justified on the basis of three independent sets of observations, of which the CMB is one. The others are the observation that the universe is expanding, first shown by Edwin Hubble in 1929 \cite{h29} and characterised by the Hubble constant, and the third is the relative abundance of light elements in the universe, such as Hydrogen and Helium \cite{t00}. 

With the Big Bang confirmed as the correct model of our universe, the next step was to start using CMB observations to better understand this model and to probe the underlying physics of both the Big Bang and the processes that led to the formation of the stars and galaxies we see today. This was the birth of CMB astronomy, which, 44 years later, culminated in the launch of the {\em Planck} mission, the European Space Agency's (ESA) CMB probe.

\section{A Very Brief History of the Universe}

\subsection{The Shape of Space}

In the 1920s, astronomers including Vesto Slipher and Edwin Hubble discovered the expansion of the universe by measuring the velocities of galaxies relative to our own. They found that, in the relatively local universe, they were moving away from us at speeds that were directly proportional to their distance, with the famous Hubble constant, $H_0$, as the constant of proportionality. Thus:
\begin{equation}
v = H_0 d
\end{equation}
where $v$ is the recession velocity of a galaxy at a distance $d$ away, and $H_0$ is the Hubble constant, measured in km/s/Mpc. These recession velocities are measured using the shift of spectral lines to the red that results from the galaxies moving away from us. This spectral shift, known as the redshift, $z$, is then:
\begin{equation}
z = \frac{v}{c} = \frac{H_0 d}{c}
\end{equation}

Within the standard Friedmann-Robertson-Walker cosmology, the expansion of the universe can be described by a single scale factor, $a$, which is defined to have a value of 1 at the present time. At earlier times $a <1$, since the universe was smaller, while in the future, since the universe will continue to expand, $a$ will be greater than 1. The rate of change of the scale factor describes the rate at which the universe is expanding, and can vary with time. The Hubble constant, $H_0$, quantifies the expansion rate of the universe at the current time, but different epochs at different times may see a different value of what is more correctly described as the Hubble parameter, $H(t)$. This is related to the scale factor $a$:
\begin{equation} 
H(t) = \frac{\dot{a}}{a}
\end{equation}
where the dot indicates differentiation with respect to time.

The wavelength of photons is also stretched by the scale factor, so that $\lambda_0 = \lambda/a$, where $\lambda$ is the emitted wavelength of a photon when the universe had a scale factor of $a$, and $\lambda_0$ is the wavelength at which we observe that photon at the present time, when $a=1$. This is what gives rise to the cosmological redshift $z$ where:
\begin{equation}
z = \frac{\lambda_0 - \lambda}{\lambda} \implies 1+z = \frac{1}{a}
\end{equation}

The dynamics of the expanding universe are governed by Einstein's General Theory of Relativity (GR), which can be applied to the universe as a whole. This yields the Friedmann equations:
\begin{equation}
\left( \frac{\dot{a}}{a} \right)^2 = \frac{8 \pi G}{3} \rho - \frac{k}{a^2} + \frac{1}{3} \Lambda
\end{equation}
and
\begin{equation}
\frac{\ddot{a}}{a} = -\frac{4 \pi G}{3} \left ( \rho + 3p/c^2 \right ) + \frac{1}{3} \Lambda
\end{equation}
and the conservation equation:
\begin{equation}
\frac{d\rho}{dt} + 3 \left( \rho + \frac{p}{c^2}\right ) \frac{\dot{a}}{a} = 0
\end{equation}
which govern the behaviour of the universe as a whole. In these equations $a$ is the scale factor, $\rho$ is the matter density, $p$ is the pressure, $\Lambda$ is the dark energy or cosmological constant term, $G$ is the gravitational constant and $k$ gives the curvature of space.

\begin{figure}
\centering
\resizebox*{10cm}{!}{\includegraphics{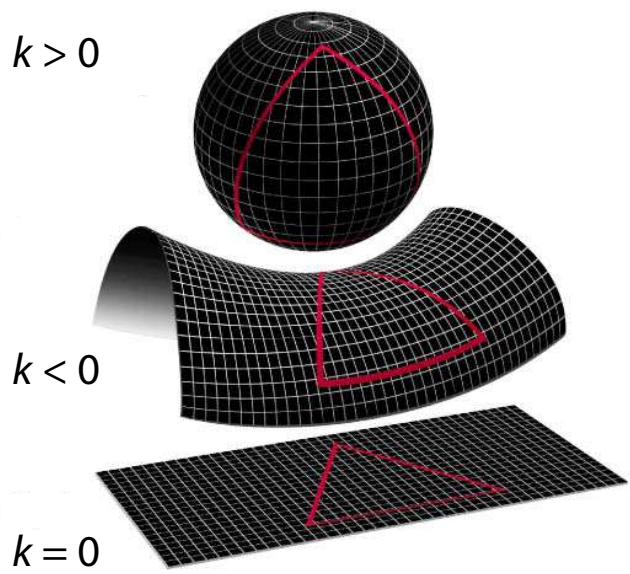}}\hspace{5pt}
\caption{The three different geometries of the universe as they would appear in 2 dimensions. Modified from an original diagram courtesy of NASA.}
\label{fig:geom}
\end{figure}

Several of these terms may be unfamiliar. The first of these is $\Lambda$. This was originally introduced into Einstein's GR equations to halt the collapse of the universe since, when they were first devised, it was thought that the universe was static. It is said that shortly after Hubble discovered the expansion of the universe, Einstein described the cosmological constant as his `greatest mistake'. More recently, it was found \cite{p98} that the expansion rate of the universe is increasing. The only way this can be accounted for in the Friedmann equations is to rehabilitate the cosmological constant in a slightly different way that allows it to drive this accelerating expansion. The physics behind this $\Lambda$ term is still unclear, and is a vigorous area of research.

The other unfamiliar term is $k$. This represents the curvature of the universe. If $k > 0$ the universe has positive curvature and, in the absence of a $\Lambda$ term, is closed, destined to expand to some maximum value and then fall back on itself into a `big crunch' at some time in the distant future. If $k < 0$ then the universe has negative curvature and will expand forever. If $k = 0$ then the geometry of the universe is precisely flat. In this case, in the absence of $\Lambda$, the universe will still expand to infinite size, but the expansion rate asymptotically approaches zero - it's like a space rocket launched from the Earth that has a velocity exactly matching escape velocity. The different geometries as they would appear in 2 dimensions are shown in Figure \ref{fig:geom}.

The value of $k$ is determined by the density of the universe - the more mass and energy in the universe the greater the curvature. For $k$ to equal zero, and the universe to be geometrically flat, the density has to have a particular value, known as the critical density. Cosmologists parameterise this using the term $\Omega$, with $\Omega = 1$ corresponding to the critical density, and thus a flat universe. Everything in the universe contributes to $\Omega$. The fraction of critical density contributed by the cosmological constant is given as $\Omega_{\Lambda}$, while the contribution of matter is given as $\Omega_m$. This term itself is divided up into two separate parts, with $\Omega_b$ representing the density of normal, baryonic, matter - the kind of material that we are made of -, and $\Omega_c$ giving the contribution of dark, non-baryonic, matter, which makes up most of the matter density of the universe but which interacts only very weakly with normal matter. $\Omega_m$ is thus the sum of $\Omega_b$ and $\Omega_c$, and
\begin{equation}
\Omega = \Omega_m + \Omega_{\Lambda} = \Omega_b + \Omega_c + \Omega_{\Lambda}
\end{equation}

The values of $k$, $\Omega_{\Lambda}$, $ \Omega_b$ and $\Omega_c$ are important factors that drive the evolution of the universe and everything in it.  Other important factors include the Hubble parameter, and the age of the universe. It turns out that observations of the CMB can, in conjunction with other data, determine the values of these parameters, and more, to unprecedented accuracy. A fuller discussion of these terms, of the Freidmann equation, and of the rest of the material in Section 2, without a number of simplifications, can be found in most introductory cosmology textbooks.

\subsection{The Emergence of the CMB}

\begin{figure}
\centering
\resizebox*{14cm}{!}{\includegraphics{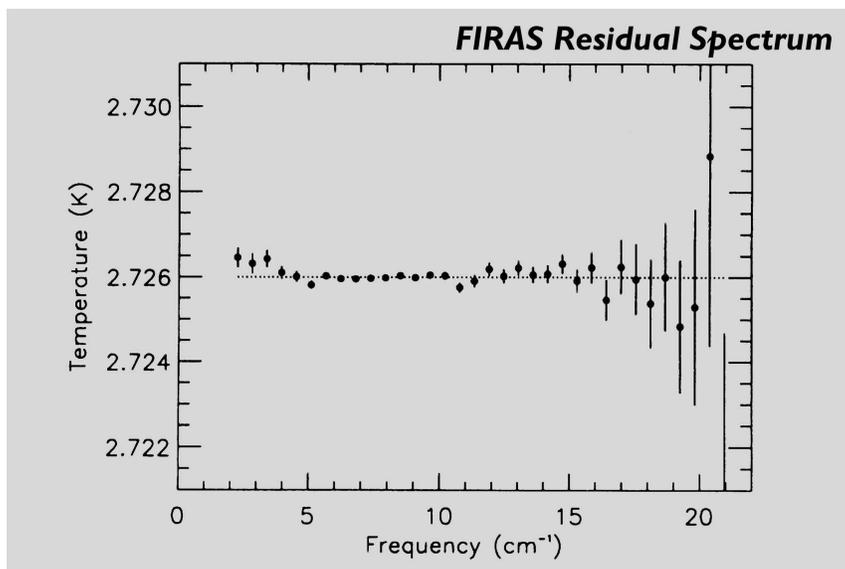}}\hspace{5pt}
\caption{Residuals of the CMB spectrum after a black body of temperature 2.726 K is removed. Courtesy of NASA.}
\label{fig:cobe_spec}
\end{figure}

To understand the origin of the CMB we must wind history backwards from what we see today to the state of the universe just a few hundred thousand years after the Big Bang. The CMB as we see it today has one of the most perfect black body spectra ever measured, very closely matching the spectral shape of a 2.726 K black body given by Planck's equation:
\begin{equation}
I(\nu, T) d\nu = \left(\frac{2 h \nu^3}{c^2} \right) \left( \frac{1}{e^{\frac{h\nu}{kT}} -1} \right ) d\nu ~ {\rm W m^{-2} ~ Hz^{-1} ~ sr^{-1}}
\end{equation}
where $\nu$ is frequency, $T$ temperature, $c$ the speed of light, $h$ Planck's constant, and $k$ Boltzmann's constant.
The best measurement of the CMB spectrum to date was made by the Cosmic Background Explorer ({\em COBE}) satellite in the early 1990s and is plotted in Fig. \ref{fig:cobe_spec}. The spectrum is such a good match to a black body that this figure shows the residual left over when a theoretical black body spectrum is subtracted from the data.

\begin{figure}
\centering
\resizebox*{14cm}{!}{\includegraphics{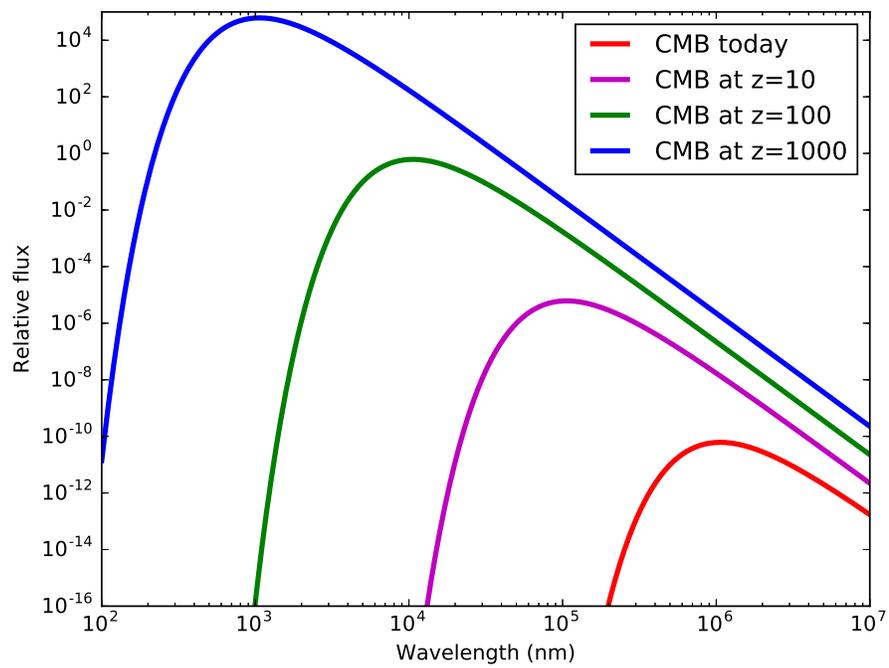}}\hspace{5pt}
\caption{The spectral energy distribution of the CMB at different redshifts, from now, at $z$=0 up to $z=1000$ where it originated \cite{m94}.}
\label{fig:cmb_z}
\end{figure}

To look back in time and see what the CMB was like at earlier epochs, we need to see what happens to a black body spectrum as both the energy density and the wavelengths of its constituent photons are modified by the scale factor, $a(t)$. If we assume that a CMB photon is emitted at some epoch $t$ with a scale factor $a$ and that it is detected at the present time, $t_0$, then the observed wavelength $\lambda_0$ is related to the emitted wavelength $\lambda$ by:
\begin{equation}
\frac{\lambda_0}{\lambda} = \frac{1}{a}
\end{equation}
and since $\lambda \nu = c $ the observed and emitted frequencies are:
\begin{equation}
\frac{\nu}{\nu_0} = \frac{1}{a} \implies \nu_0 = a \nu
\end{equation}
Similarly it can easily be shown that:
\begin{equation}
d\nu_0 = a~d\nu
\end{equation}
The energy density of a black body spectrum is given by:
\begin{equation}
U(\nu, T) d\nu = \left(\frac{8 \pi h \nu^3}{c^3} \right) \left( \frac{1}{e^{\frac{h\nu}{kT}} -1} \right ) d\nu~{\rm J~m^{-3}~Hz^{-1}}
\end{equation}
so the number of photons per unit volume between frequencies $\nu$ and $\nu + d\nu$, and thus with energy $h\nu$, is:
\begin{equation}
n(\nu, T) d\nu = \left(\frac{8 \pi \nu^2}{c^3} \right) \left( \frac{1}{e^{\frac{h\nu}{kT}} -1} \right ) d\nu~{\rm m^{-3}~Hz^{-1}}
\end{equation}
Photons emitted between $\nu$ and $\nu + d\nu$ will be observed between $\nu_0$ and $\nu_0 + d\nu_0$, while all volumes increase by a factor of $1/a^3$. This means that the number density of photons with frequencies between $\nu_0$ and $\nu_0 + d\nu_0$ becomes:
\begin{equation}
n(\nu_0,T)d\nu_0 = a^3 \left( \frac{8\pi}{c^3} \right) \left(\frac{\nu_0}{a}\right)^2 \left( \frac{1}{e^{\frac{h\nu_0}{kaT}} -1} \right ) \frac{d\nu_0}{a} = \left( \frac{8\pi \nu_0^2}{c^3} \right)  \left( \frac{1}{e^{\frac{h\nu_0}{kaT}} -1} \right ) d\nu_0~{\rm m^{-3}~Hz^{-1}}
\end{equation}
We multiply this by $h\nu_0$ to get the energy density of this observed spectrum, and that gives us:
\begin{equation}
U(\nu_0, T) d\nu_0 = \left(\frac{8 \pi h \nu_0^3}{c^3} \right) \left( \frac{1}{e^{\frac{h\nu_0}{kaT}} -1} \right ) d\nu~{\rm J~m^{-3}~Hz^{-1}}
\end{equation}
which is simply a black body spectrum with a temperature of $T_0 = a T$. The radiation spectrum is thus still a black body, but at a temperature multiplied by the scale factor. Therefore, as we look back in time to epochs with a smaller scale factor, the CMB will remain a black body, but at a higher temperature given by $T= 2.726 / a$, or, alternatively, $T=2.726 (1+z)$ (see Fig. \ref{fig:cmb_z}).

\subsection{The Surface of Last Scattering}

The photons that make up the CMB are not the only things in the universe. There is also matter. Most of this is made up of the still poorly-understood dark matter, but some is made up of baryonic matter. This is the matter we are used to, made up of protons, neutrons and electrons. Today, these particles are bound into a range of different elements - including carbon, oxygen, iron, all the way up to uranium. Almost all of these elements were cooked up in stars. In the early universe, at the time the CMB was produced, the abundance of elements was dominated by hydrogen, with about 25\% by mass in the form of helium. There were also trace amounts of deuterium and lithium, which are important for understanding the details of the Big Bang, but which are not significant for our purposes here.

There are about $10^9$ photons for each baryon in the universe, a figure which can be calculated by comparing the number density of photons in the CMB to the number density of protons and neutrons in matter. At the early stages of the universe, as we have seen, the CMB will be much hotter than it is today. Go back far enough and the CMB will contain enough high energy photons to ionise every hydrogen and helium atom in the universe. This means that the original state of baryonic matter in the universe is a plasma, with electrons, protons and some helium nuclei, all dissociated and moving freely.

Radiation interacts far more strongly with a plasma than with neutral, un-ionized matter, so the photons and baryons are tightly coupled, with the universe being effectively opaque to photons. As the universe expanded, the black body spectrum of the CMB dropped in temperature until the point is reached when there are too few high energy photons to keep the hydrogen and helium in the plasma ionised. This means that electrons and protons will bind together to produce hydrogen atoms, and similarly for helium. This is known as the epoch of recombination  - something of a misnomer since the electrons and nuclei combining together to form neutral atoms had never before been combined. After recombination the universe is no longer filled with a plasma. Instead it is filled with neutral atoms which no longer interact strongly with the photons of the CMB. The universe has become transparent, with photons now able to free stream in all directions.

\begin{figure}
\centering
\resizebox*{10cm}{!}{\includegraphics{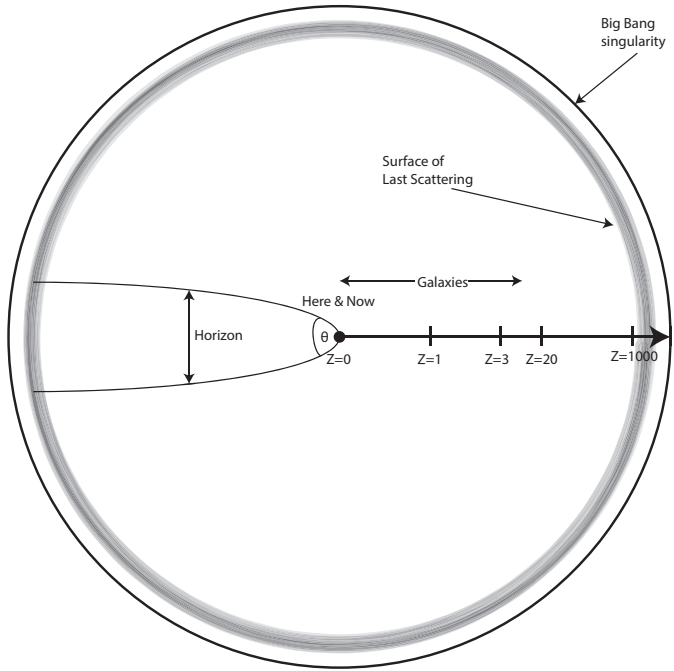}}\hspace{5pt}
\caption{Our view of the universe centred on the observer. The angle represents the angle of view and the radius represents both the distance and time of an event, since light travels at a finite speed. The distance scale is non-linear and marked in terms of redshift. The surface of last scattering is the origin of CMB photons at a redshift of 1000.}
\label{fig:sols}
\end{figure}

These are the photons that make up the CMB. They have been travelling across the universe since the time of recombination, largely unimpeded by interactions with matter since they last scattered off the hot plasma of the early universe. The CMB is thus a record of what the universe was like at this time, and what we see when we observe it is a view of what is called the `surface of last scattering' (see Fig. \ref{fig:sols}).

Most CMB photons will never have the misfortune to be trapped by a cosmologist's telescope and absorbed by an instrument. But the tiny fraction that are captured allow us to see what the universe was like at the epoch of recombination, just $\sim$400,000 years after the Big Bang.

\subsection{CMB Anisotropies} 

Since baryonic matter is tightly coupled to photons in the plasma that filled the universe before recombination, any differences in the density of baryonic matter from one place to another will be reflected in temperature differences in the CMB - where there is more baryonic matter there will be a higher temperature. We expect there to be density and thus temperature variations in the CMB because of the large scale structures that we see in the universe today, from galaxies, to galaxy clusters, superclusters and beyond. These structures have all grown through gravitational attraction from seed density fluctuations at a level of about one part in $10^5$. The origin of these seed fluctuations likely arises at much earlier stages of the universe, during what is known as the inflationary epoch \cite{infl} \cite{l99}, and from physics that is not yet well understood. However, the fluctuations must be there, otherwise the universe today would be very different, and we would not exist.

\begin{figure}
\centering
\resizebox*{14cm}{!}{\includegraphics{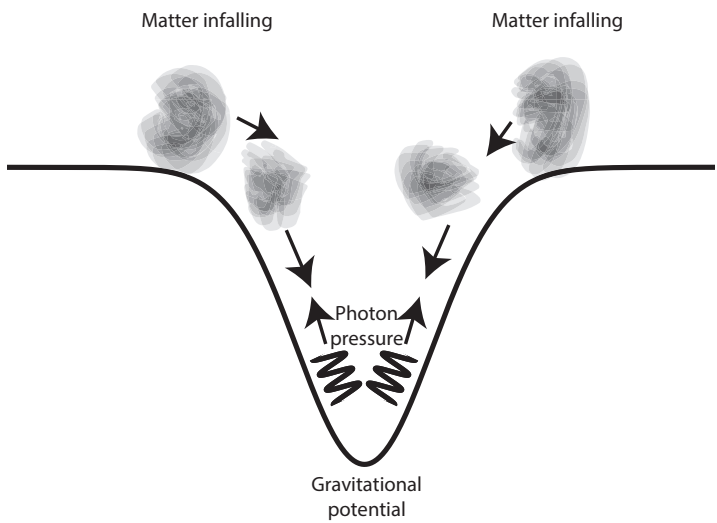}}\hspace{5pt}
\caption{Matter falling into a region of enhanced density due to gravity is pushed back by photon pressure, leading to acoustic oscillations.}
\label{fig:oscill}
\end{figure}

\begin{figure}
\centering
\resizebox*{12cm}{!}{\includegraphics{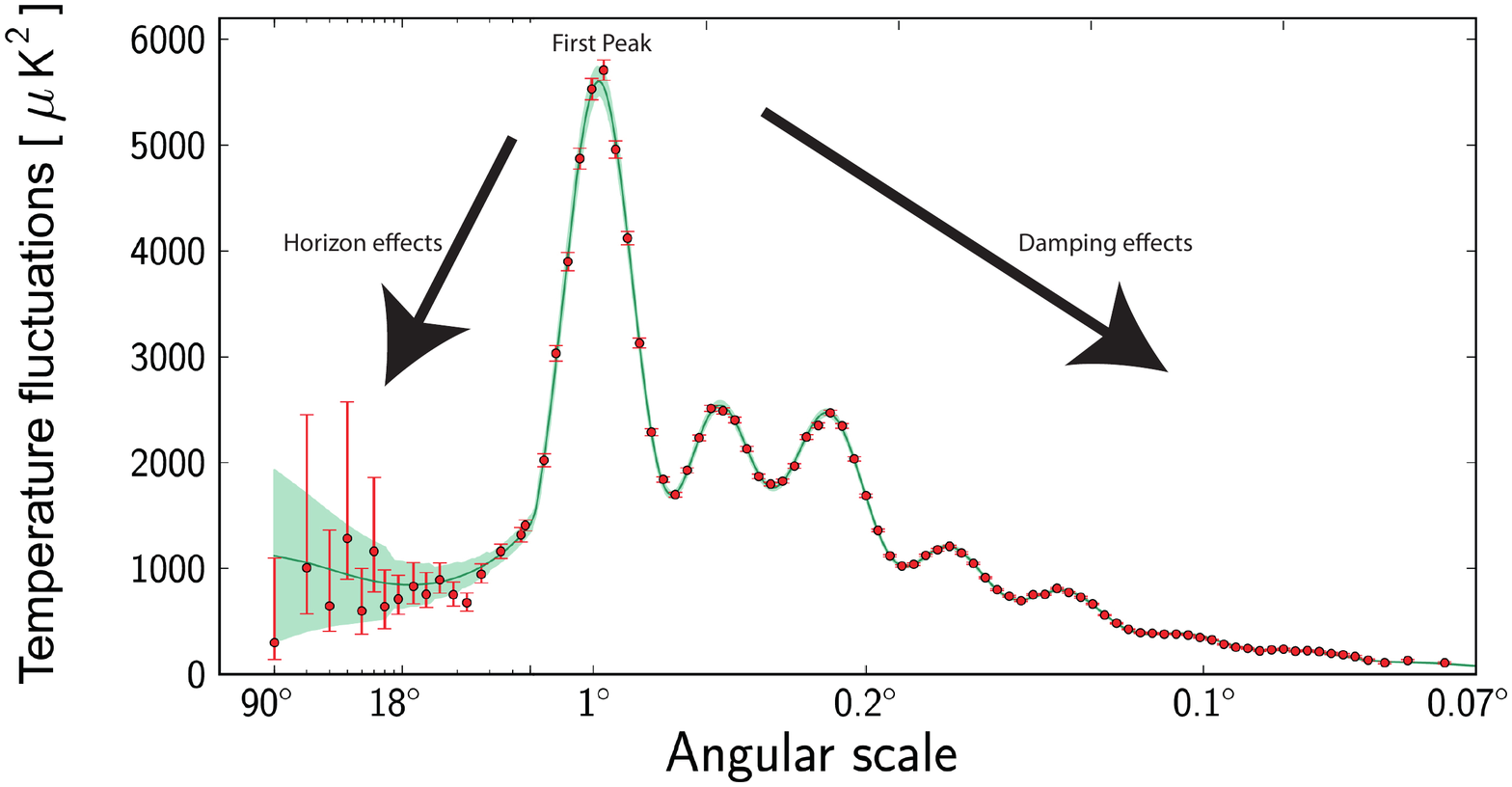}}\hspace{5pt}
\caption{The effects of damping and the horizon on acoustic oscillations at the surface of last scattering lead to a peak in the power spectrum of anisotropies. This figure shows both a theoretical model for the anisotropy spectrum (solid line) and data from the {\em Planck} satellite. As you can see the model and data are almost indistinguishable. From {\em Infrared Astronomy: Seeing the Heat} - used with permission.}
\label{fig:planck_ps}
\end{figure}

The initial density fluctuations will be present in both baryonic and dark matter. Since dark matter particles do not interact with photons, the dark matter density fluctuations are free to grow gravitationally. Interactions between baryonic density fluctuations and the photons that eventually become the CMB, though, produce something rather different. Baryons falling into a region of overdensity through gravity will meet an enhanced number density of photons which, through Thomson scattering, will exert a force in the opposite direction (see Fig. \ref{fig:oscill}). The result of this is that the baryons undergo acoustic oscillations, alternately falling towards a region of enhanced density and then being pushed away by photon pressure. The largest physical size these oscillations can have can be shown to be about 100 kpc in size. Smaller scale structures are suppressed by photons diffusing out of them in a process known as Silk Damping \cite{s68}, while larger scale structures cannot grow because the universe is not old enough for light to travel between them. This means that they are causally disconnected and thus cannot affect each other (see Fig. \ref{fig:planck_ps}). These two effects combine to produce a characteristic scale for the strongest CMB anisotropies. This is about a degree in size on the sky. The exact scale and strength of this peak, and the strength and sizes of the smaller scale peaks in the anisotropy power spectrum, depend on the precise values of a range of cosmological parameters, including the curvature of the universe ($k$), the contributions of baryonic matter, dark matter and dark energy to its constituents, the age of the universe, and a number of other values that give a precise description of our universe. Detecting and measuring the properties of CMB anisotropies thus became a central goal of observational cosmology very soon after the CMB was discovered.

\section{Observing the CMB}

Measuring CMB anisotropies requires precise observations using very sensitive detectors so that temperature differences smaller than 1 part in $10^5$ can be determined. Needless to say, this is a difficult task, not helped by the fact that everything in the universe, from the Earth's atmosphere to distant galaxies, lies between us and the surface of last scattering. These `foregrounds' can completely swamp the CMB anisotropy signal, and so they must be very accurately measured before they can be removed, and the cosmological signal is revealed. It is therefore not surprising that it was more than 25 years after the discovery of the CMB before the first cosmological anisotropy signal was detected.

The first deviation from uniformity in the temperature of the CMB was discovered in the 1970s by airborne observations using balloons \cite{c76} \cite{h71} or a high altitude U2 aeroplane \cite{s77}, to escape some of the backgrounds coming from the Earth's atmosphere and from man-made sources. What they discovered were not cosmological anisotropies, but the effect of the Earth's motion relative to the CMB. This produces a Doppler effect, boosting the energy of the photons coming towards us from the direction the Earth is travelling in, and reducing the energy of photons coming from the opposite direction. The overall effect means that the CMB is slightly warmer in our direction of travel and slightly cooler in the opposite direction. This produces a dipole distortion in the CMB with an amplitude of 3.4mK corresponding to a velocity of 270 km/s.

\begin{figure}
\centering
\resizebox*{13cm}{!}{\includegraphics{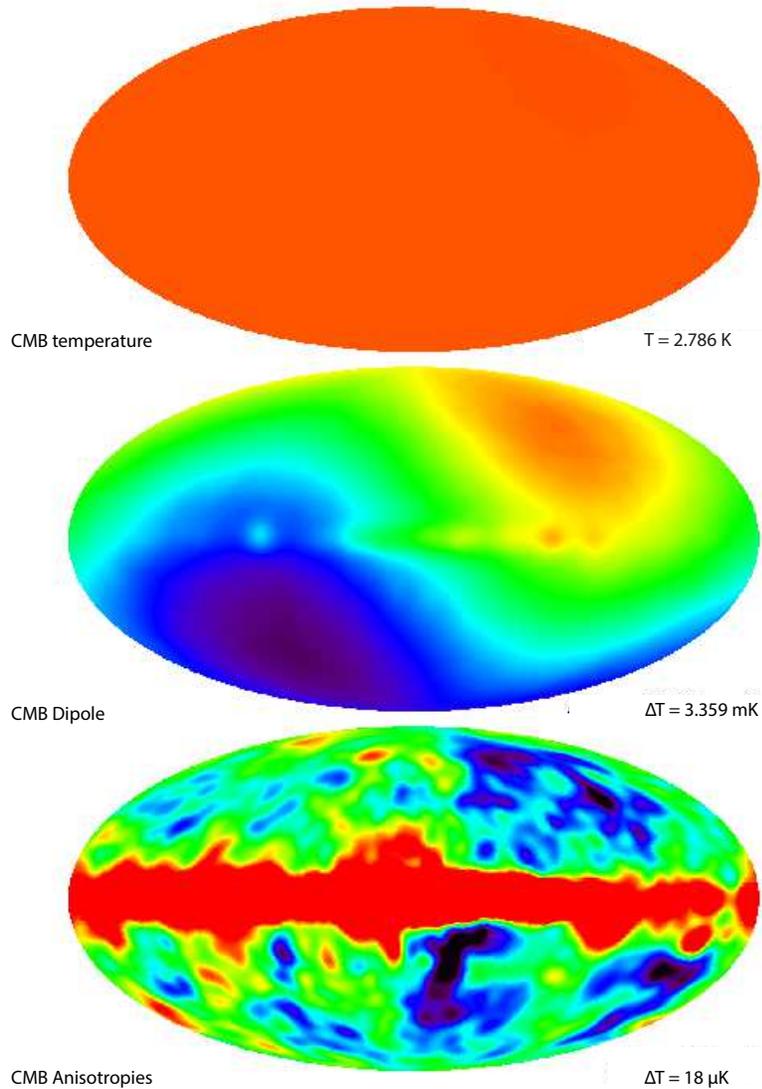}}\hspace{5pt}
\caption{The CMB as viewed by the {\em COBE} satellite, plotted in galactic coordinates using a Mollweide projection so they cover the entire sky. The top image shows the temperature of the CMB over entire sky as observed by {\em COBE}. As you can see it is highly uniform. If you subtract off the mean temperature and the plot the temperature variations across the entire sky you get the middle image. This shows the CMB dipole. This has to be removed to reveal the cosmological anisotropies, which can be seen in the bottom image. In the two lower images you can also see a band across their middle. This band of emission comes from emission in the plane of our own galaxy and thus is nothing to do with the CMB. Courtesy of NASA. }
\label{fig:cobe_cmb}
\end{figure}

The use of balloons and a U2 spy plane for the dipole observations were the first step in CMB astronomers' search for the best place to conduct observations. While the initial discovery of the CMB took place in the unexceptional environment of New Jersey, the need to remove all possible contaminating emission has led CMB astronomers to the tops of mountains, to high altitude deserts, to polar icecaps, and to the edges of space with balloons and aircraft. However, it was not until the hunt for primordial anisotropies moved to space that the first detections were made. These came in the early 1990s with NASA's {\em COBE} satellite \cite{s92}. The amplitude of these fluctuations was just 30$\mu$K, over a hundred times weaker than the dipole anisotropy which itself was a thousand times weaker than the $\sim$3K background temperature itself. Images of the CMB sky in temperature,  with the mean value removed to reveal the dipole, and with the dipole removed to reveal the cosmological anisotropies, all from the {\em COBE} mission, are shown in Fig. \ref{fig:cobe_cmb}.

The cosmological anisotropies uncovered by {\em COBE} are on scales of ten degrees and larger. This means that they are on much larger scales than those predicted for the first peak in the CMB power spectrum. The instrumentation on board {\em COBE}, and in particular the Diffuse Microwave Radiometer (DMR) which produced these results, had an angular resolution of only seven degrees, so smaller structures could not be detected. The announcement of the {\em COBE} results added to the impetus of other projects looking for CMB anisotropies on smaller scales. A new generation of balloon-borne CMB telescopes was launched, and a new space mission, the Wilkinson Microwave Anisotropy Probe ({\em WMAP}), began construction.

The balloon missions, such as BOOMERanG \cite{m00}, got to take data first, and were the first to observe the peak in the power spectrum. Balloon observations, though, can only cover a small subregion of the sky. To cover the entire sky requires a space mission like {\em WMAP}, which not only confirmed the results of the balloon missions but allowed for much more detailed studies of the statistics of the anisotropies. The best, however, was yet to come.

\section{The {\em Planck} Mission}

In the early 1990s, CMB research groups across Europe were preparing proposals for a European Space Agency (ESA) mission that would go far beyond the capabilities of {\em COBE}, which had just announced the first detection of CMB anisotropies. Two separate missions emerged: {\em COBRAS} (the Cosmic Background Radiation Astronomy Satellite) which would work at high frequency radio wavelengths and conduct observations on angular scales down to 0.5 degrees - largely matching the capabilities of the NASA {\em WMAP} mission; and {\em SAMBA} (the Satellite for Measurement of Background Anisotropies) a significantly more ambitions project that would work to much smaller angular scales, about 5 arcminutes, by observing at higher frequency far-infrared and millimetre wavelengths. Both proposals were assessed and ESA decided that an ideal CMB mission could be produced by combining the two satellites together into what would become the {\em Planck} mission.

\subsection{The Need for Planck}

The central goal for the {\em Planck} mission was, as the mission's scientific programme book put it, to `extract essentially all of the information in the CMB temperature anisotropies' \cite{pl06}. This means that {\em Planck} would effectively mark the end point for all the work on temperature anisotropies that had begun with the discovery of the CMB in 1964. This would allow cosmological parameters to be determined to much higher accuracy than ever before, measuring such things as the age of the universe, the amount of baryonic matter, dark matter, and dark energy to precisions of a few percent.

However, CMB observations do not end with measurements of temperature. {\em Planck}'s other main cosmological goal, after finishing off the field that had occupied the lives of CMB astronomers for over 40 years, would be to advance the next phase of background radiation studies by conducting sensitive observations of the polarization of the CMB, something that was first attempted on large scales by {\em WMAP}. Polarization is the next step for CMB observations since it potentially holds the key to physics at much earlier stages of the universe than can be accessed from temperature anisotropies. A particular pattern of polarization in the CMB, known as B-mode polarization \cite{h97}, can only be produced by gravitational waves. While gravitational waves are produced in the universe today by physical processes involving compact stars and black holes, they can only arise in the early universe from physics associated with inflation \cite{k16}. This is a process suspected to take place in the very first instants of the Big Bang, since it solves a variety of problems with the standard picture of cosmology \cite{l99}. However, direct evidence for inflation has yet to be been found. The physics that drives inflation would go beyond the standard model of particle physics, and so would provide clues to the ultimate physical theory that would unify quantum mechanics, the physics of the very small, with general relativity, the physics of gravity and the universe as a whole.

To achieve these goals the {\em Planck} mission would also have to determine the contributions from astrophysical foregrounds to very high precision, so that they can be removed. These astrophysical foregrounds come from a wide variety of sources spread throughout the universe, from dust in our own Solar System to the radio emission of the most distant quasars. There is thus a wealth of astrophysics within this foreground data. {\em Planck}'s science goals therefore also include studies of galaxy clusters, radio galaxies, dusty galaxies, dust, gas, radio emission and magnetic fields in our own Galaxy, and the study of dust in the Solar System \cite{pl06}.

\subsection{Mission Concept}

These science goals set a number of requirements for the design of the {\em Planck} spacecraft, instruments, and mission. They include:

\begin{itemize}

\item Measurement of CMB fluctuations down to angular scales as small as 5 arcminutes; this means that the primary mirror of the instrument has to be comparatively large. If we assume the instrument will be working at the diffraction limit the angular size of the beam is $1.22 \lambda/D$, where $\lambda$ is the wavelength of observation and $D$ is the diameter of the mirror. Operation at the peak wavelength of the CMB, at $\sim$1.4 mm, means that the mirror will have to be 1.2 m in diameter or larger. This is quite a large mirror to put on a space telescope - the Hubble Space Telescope mirror is only 2.5 m in diameter - so this sets strong constraints on the design of the mission.

\item Very sensitive measurement of both intensity and polarisation anisotropies. Temperature accuracies in terms of $\Delta T/T$ of around 2.5$\times 10^{-6}$ are needed. This sets strong requirements on the sensitivity and thus the design of the instruments. Following the COBRAS/SAMBA heritage, two instruments working with different technologies and at different wavelength regimes are used.

\item Scan the entire sky several times over the course of a mission that will last at least 18 months. Combining this with the need to have the detectors kept at very low temperatures to reach the required sensitivity means that an orbit close to the Earth, where the spacecraft is continuously bathed in both direct sunlight, reflected sunlight from the Earth, and Earth's own thermal radiation, is not a feasible option. 

\item The ability to measure and remove a wide range of different astrophysical foregrounds so that the underlying CMB anisotropies can be measured to greater precision, and so that a wide range of foreground studies can be undertaken. The use of two instruments, which were named the Low Frequency Instrument (LFI) and the High Frequency Instrument (HFI), covering nine different frequency bands from 30 GHz to 857 GHz (1cm to 350$\mu$m  in wavelength) makes this possible, as we will see below.

\end{itemize}

Needless to say there were many other requirements on the design of {\em Planck}, but the above set the main constraints for the mission that was eventually flown. 

\section{The Planck Spacecraft}

\begin{figure}
\centering
\resizebox*{6cm}{!}{\includegraphics{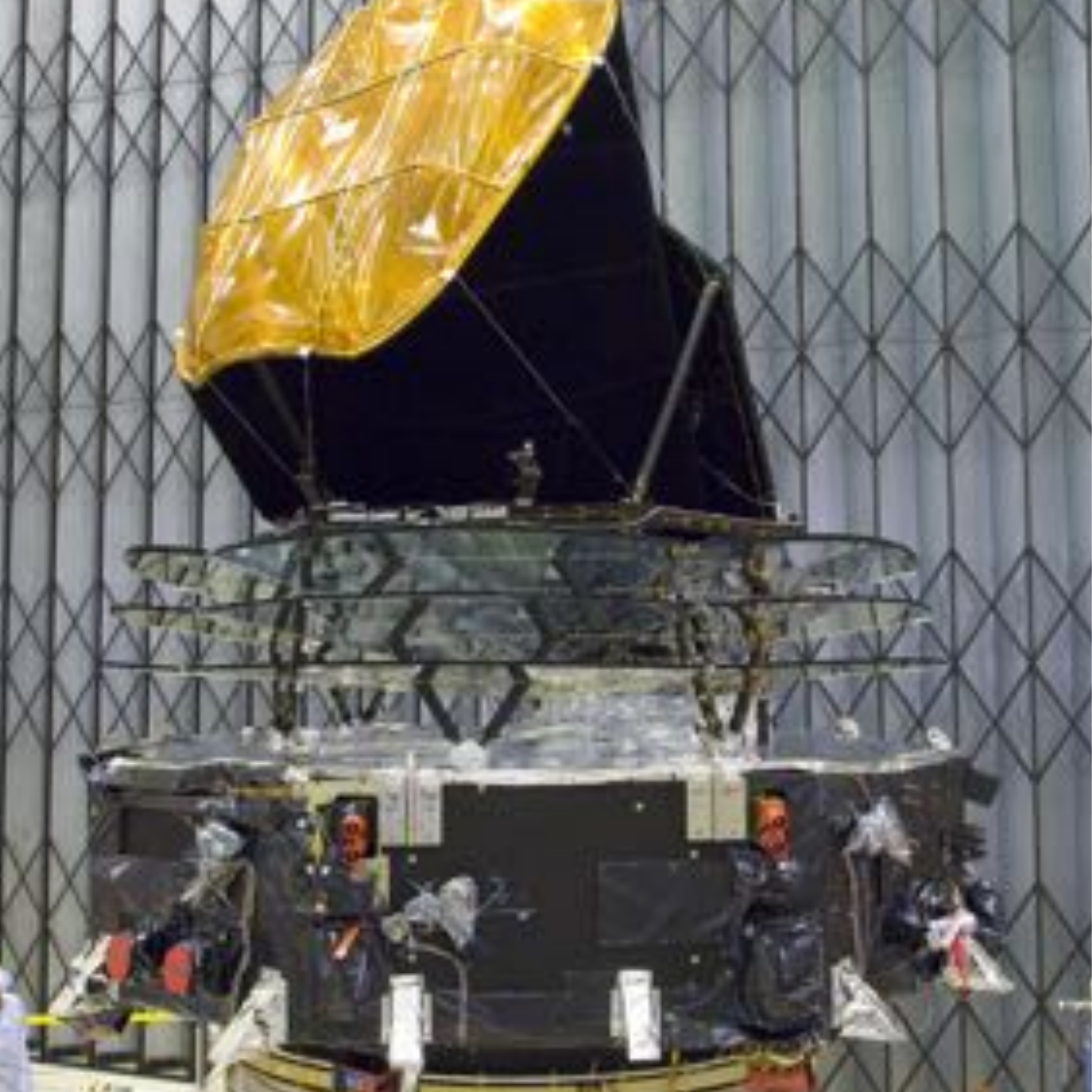}}\hspace{5pt}
~\\~\\
\resizebox*{12cm}{!}{\includegraphics{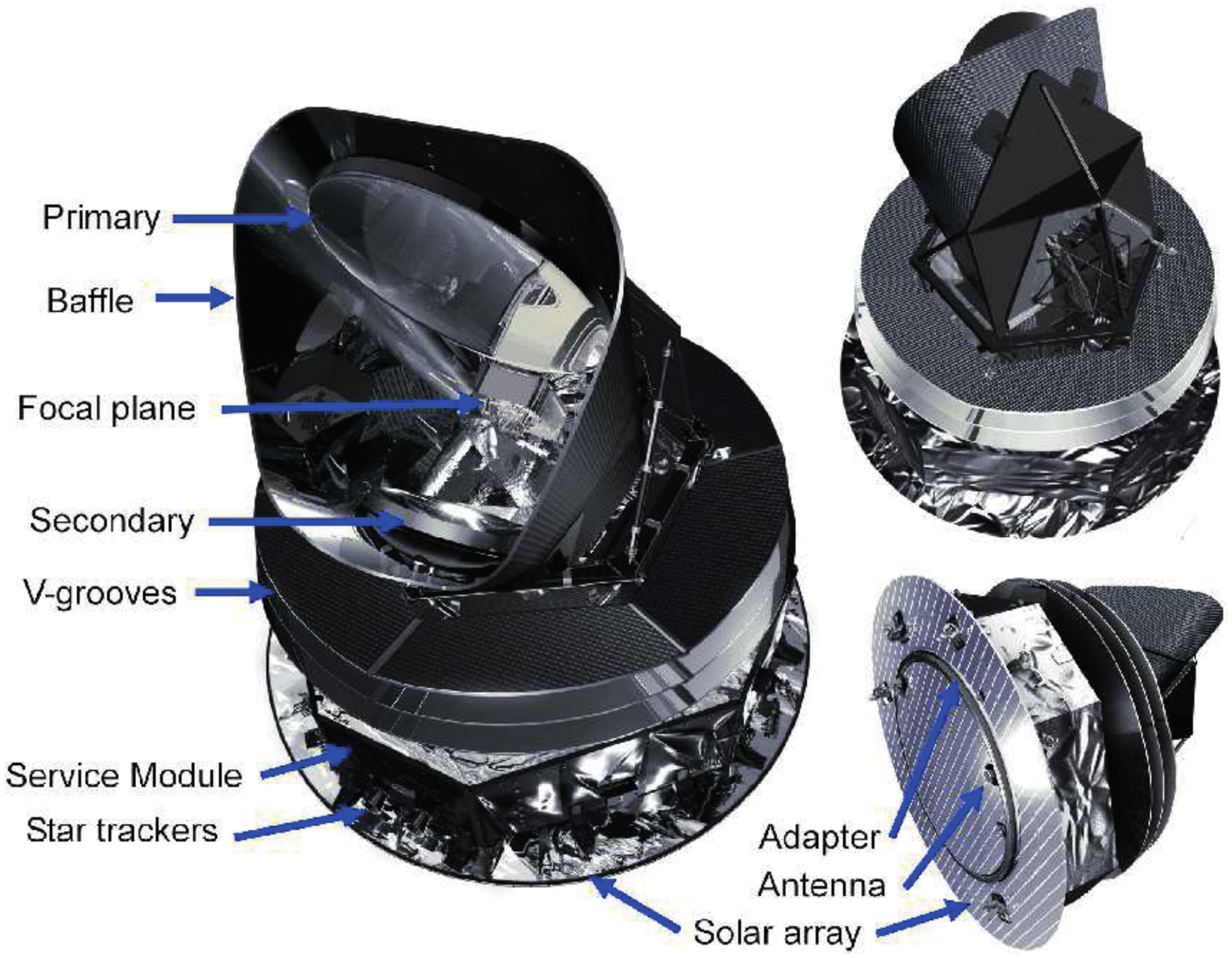}}\hspace{5pt}
\caption{The {\em Planck} spacecraft. Top: during testing on the ground, with the mirrors covered by protective sheets. Bottom: in artist's conception showing various elements of the design. Both images courtesy of ESA.}
\label{fig:planck_craft}
\end{figure}

The {\em Planck} spacecraft is shown in Fig \ref{fig:planck_craft}. The primary mirror is an off axis reflector with a projected diameter of 1.5 m, allowing for high resolution studies of the CMB. The mirror and focal plane instruments are protected from scattered light by a large baffle. A series of V-grooves further reflect away incident radiation from the instruments and mirror, and thermally isolate them from the service module that makes up the rest of the spacecraft. The service module contains a series of refrigeration systems that keep the instruments cool as well as the warm electronics to run the instruments, and control the spacecraft.

The HFI and LFI instruments are mounted in the focal plane and are made up of multiple feed horns that deliver radiation to single pixel detectors operating at the appropriate frequencies. The spacecraft rotates at a rate of one revolution per minute allowing it to scan the detectors around a circle on the sky every minute. The axis of rotation is then shifted about once every 50 minutes \cite{p11_1} by two arcminutes. This allows the entire sky to be surveyed over the course of six months. The full scanning strategy is discussed in \cite{taub10}.

\subsection{The Low Frequency Instrument (LFI)}

\begin{figure}
\centering
\resizebox*{12cm}{!}{\includegraphics{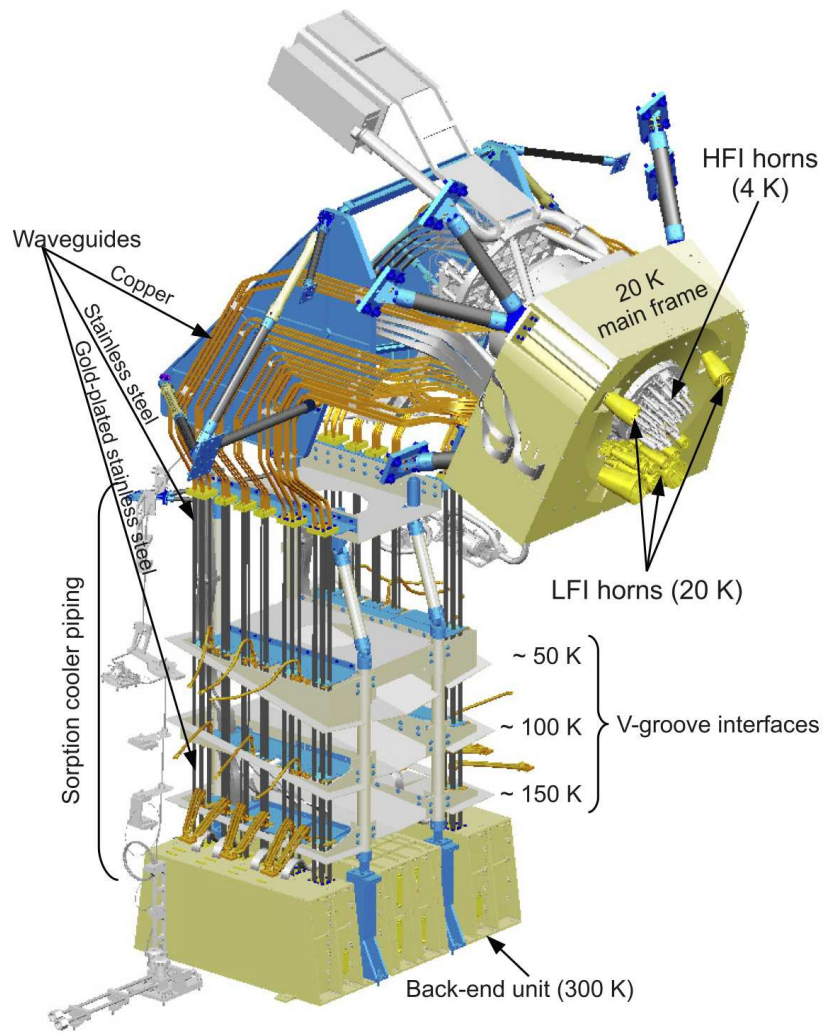}}\hspace{5pt}
\caption{The {\em Planck} LFI instrument. Left: A 3D view of LFI showing thermal stages, waveguides, cooler piping and the focal plane. The LFI is built around the HFI and uses the HFI 4K thermal stage as a reference load for its differential measurements of the flux incident from the sky. Right: The {\em Planck} focal plane showing the feed horns of both the LFI (numbered) and HFI instruments. Each of these eleven feed horns leads to a single pixel detector. From \cite{m11}.}
\label{fig:lfi}
\end{figure}

The Low Frequency Instrument (LFI, see Fig. \ref{fig:lfi}) operates at high radio frequencies, in three bands at 30, 44 and 70GHz (1cm, 0.68 cm and 0.43 cm in wavelength respectively) using high electron mobility transistors (HEMT) based on indium phosphate. These have an operating temperature of 20 K. The receivers all work in a differential mode, with the signal from the sky continuously compared to a stable 4 K reference load that is mounted on the 4 K radiation shield of the HFI instrument. This signal is then carried along waveguides that pass through the V-groves to the warm electronics in the spacecraft service module where they are amplified and recorded. Each LFI receiver operates in two orthogonal polarisations, so that the polarisation of the signal as well as its intensity can be measured. The LFI receivers are amongst the most sensitive radio receivers ever built at these frequencies.

\subsection{The High Frequency Instrument (HFI)}

\begin{figure}
\centering
\begin{tabular}{cc}
\resizebox*{7cm}{!}{\includegraphics{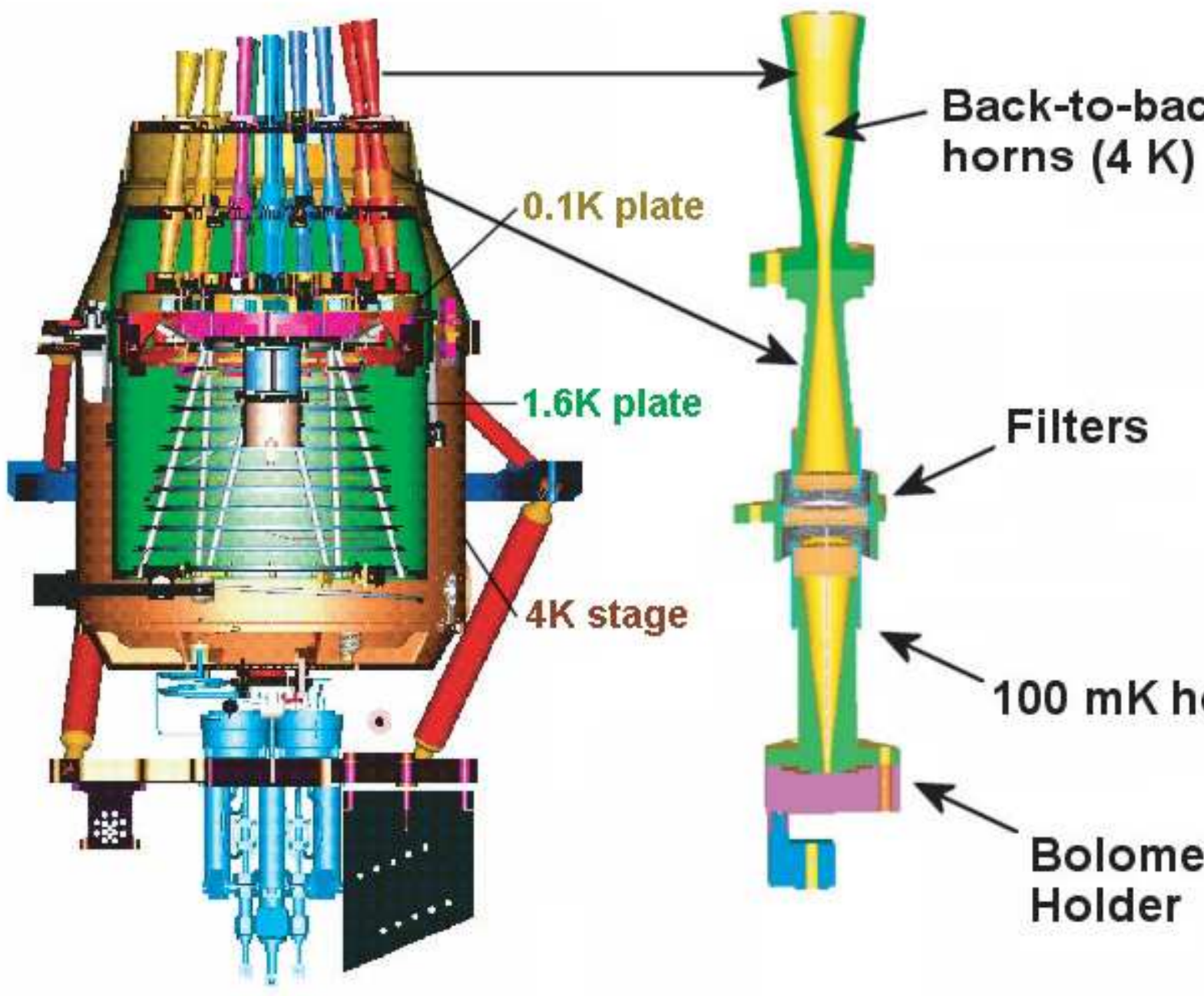}}\hspace{5pt}&
\resizebox*{7cm}{!}{\includegraphics{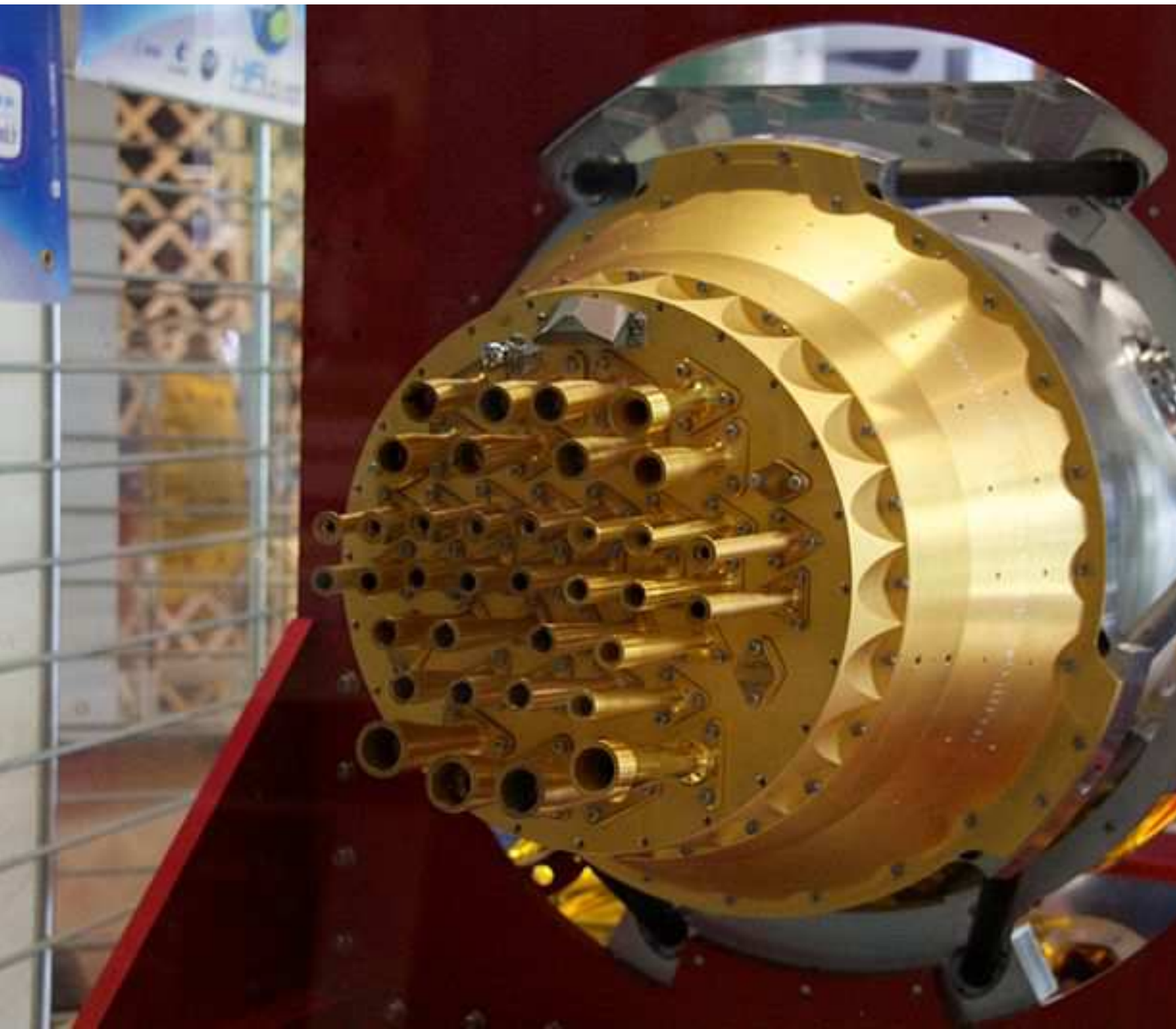}}\hspace{5pt}\\
\end{tabular}
\caption{The {\em Planck} HFI instrument. Left: A cut away view showing the various cooling stages and the design of the feed horns and filters that direct light onto the detectors. Image courtesy of ESA. Right: The assembled HFI test model, identical to the actual HFI before installation into the {\em Planck} focal plane, showing the feed horns and the 4 K radiation shield. Image courtesy of Dr Mike Peel.}
\label{fig:hfi}
\end{figure}

The High Frequency Instrument (HFI, see Fig. \ref{fig:hfi}) covers six frequency bands from 100 GHz to 857 GHz (3 mm to 350$\mu$m in wavelength) in the mm to submm part of the spectrum. The lower four frequencies include sensitivity to polarisation as well as intensity. The detectors used are silicon spiderweb bolometers. These detect absorbed radiation through measuring the small temperature change this produces. They consist of two components: a radiation absorber, and a thermistor which can measure small changes in temperature through changes in electrical resistance. To reach the required sensitivity these thermistors, and thus the bolometer devices, have to be cooled to a temperature of just 100 mK. This means that the {\em Planck} HFI was the coldest place in space while it was operating.

\subsection{Cooling System}

Operating a complex astronomical instrument in space at temperatures lower than can be obtained in most physics laboratories represented a significant technical challenge which had to be overcome if {\em Planck} was to be a success. The cooling system for {\em Planck} is thus a centrally important part of the mission.

Before {\em Planck}, most missions requiring cryogenic temperatures have used a liquid or solid gas as the main coolant. Examples of this include the {\em IRAS, ISO} and {\em Herschel} space missions, all of which had large liquid helium tanks to provide the cooling. There are two disadvantages to this approach. Firstly, the liquid helium adds to the mass and size of the spacecraft, increasing the costs of launch and construction. Secondly, the liquid helium is an expendable resource, boiling off as it cools the instruments to the required temperature. Once the helium is gone there is no more cooling, unlike the refrigerators in our kitchens which operate on a closed cycle and merely require electricity to function. 

The cooling solution eventually designed for {\em Planck} is a multistage process that shaped much of the design of the spacecraft and the selection of its orbit. Unlike most previous astronomy missions it does not use bulk liquid helium for cooling. Instead it uses passive cooling, two different closed-cycle refrigerators, and, for the lowest temperatures, an open-cycle dilution refrigerator \cite{p11_1}. In detail the cooling stages are as follows:

\begin{itemize}

\item Passive cooling

All cooling on space telescopes was done with liquid gases until the {\em Spitzer} mission, launched in 2003 \cite{w04} which combined traditional cryogenic gas and passive cooling techniques. This substantially reduced the launch mass of the spacecraft and extended its cold operational lifetime. The idea for passive cooling, originally devised by Tim Hawarden and Harley Thronson in their proposal for the {\em Edison} mission \cite{t95}, a passively cooled infrared space telescope. The idea behind passive cooling is very simple: space is cold (the CMB temperature in fact) so a telescope and set of instruments, appropriately shielded from incident radiation from the Sun and other sources of heat, gradually cools down through radiation. {\em Spitzer} was the first mission to put this into effect. It was equipped with a large Sun shield and launched into an Earth trailing orbit to avoid reflected sunlight and thermal radiation. The result was a great success and essentially every single space observatory since has used passive cooling at some level.

In the case of {\em Planck} an orbit around the L2 point (a stable point roughly 1.5 million kilometres from the Earth in the direction away from the Sun) was selected. At this position the Sun and Earth are always in the same direction. A Sun shield, that includes solar cells on the base of the spacecraft, the V-grooves, and baffling around the telescope, all kept out incident sunlight. This allowed the telescope and instruments to cool to $\sim$50K over a period of four weeks following launch.

\item Sorption Cooler

Cooling from 50K to 18K is done with adsorption cooling, where six metal alloy beds adsorb hydrogen gas acting as compressors. The system works in a similar way to a conventional domestic refrigerator where a compressed gas, driven off the sorption beds through heating, is allowed to expand. In this case liquid hydrogen droplets form on expansion to provide the cooling. While gas from one of the sorption beds is expanding, others are cooling through radiation to space, and adsorbing hydrogen for later expansion. In this way a continuous supply of cold hydrogen is provided.

\item $^4$He Joule-Thomson Cooler

This device allows cooling from 18 K to 1.4 K. It uses two mechanical compressors to alternately compress and then expand $^4$He gas. The use of $^4$He rather than hydrogen allows lower temperatures to be reached.

\item Dilution cooler

The final stage of cooling is the only part of the {\em Planck} cooling system that involves expendable material. This is the Dilution cooler. It operates by diluting pure liquid $^3He$ into a mixture with $^4He$. This allows cooling from 1.4 K to the 100 mK necessary for operation of the HFI bolometers. The pure $^3He$ and $^4He$ used in this process are stored in high pressure tanks in the spacecraft service module, and vented into space once used. When these expendables are used up the HFI can no longer function. The dilution cooler is housed inside the 4 K radiation shield at the back of the HFI focal plane (see Fig. \ref{fig:hfi} left).

\end{itemize}

\section{Launch and Operations}

{\em Planck} was launched on 14th May 2009 on an Ariane 5 ECA rocket from the ESA launch site in Kouru in French Guyana. The launch vehicle was shared with the {\em Herschel} Space Observatory which sat above {\em Planck} in the launch bus. Once the upper stage of the rocket reached Earth orbit first {\em Herschel}, and then {\em Planck} were released. {\em Planck} then conducted three orbital manoeuvres to place it in its final orbit around L2 \cite{p11_1}.

{\em Planck} reached its final orbit on 3 July 2009. During the nearly two months of travel from Earth to L2, the telescope passively cooled from room temperature to 50K. Various preparation activities were carried out, including heating of various parts of the spacecraft to allow outgassing. This decontamination phase lasted two weeks. After this, the instruments were cooled to their operating temperature with the active cooling system. These preparation activities took a total of two months, so the spacecraft was ready to begin commissioning observations when it arrived at L2. Calibration and performance tests proceeded until the end of August 2009. The first normal observations taken formed the First Look Survey (FLS), which started on 13 August 2009 as part of these tests. This was completed successfully on 27 August 2009. {\em Planck} then continuously scanned the sky, completing a survey of the entire sky roughly every 6 months.

The expendables in the HFI dilution cooler ran out after 29 months of operation. HFI observations ceased at that time, after completing roughly 4.5 surveys of the entire sky. The LFI continued to operate with just the closed cycle coolers for a total of 48 months, allowing it to survey the entire sky roughly 8 times. The original requirement for the mission was for HFI and LFI to complete just two surveys of the entire sky. The entire {\em Planck} dataset is available to the astronomical community and the general public, as well as numerous derived products such as source catalogs, CMB maps etc. They can be obtained from the {\em Planck} Legacy Archive website: \verb+http://pla.esac.esa.int/pla/+. Overall instrument performance met or exceeded the requirements of the mission (see Table \ref{tab:inst}), while the overall mission lifetime was more than twice as long as required.

\begin{table}
\begin{tabular}{cccc|cccccc}\hline
Instrument&\multicolumn{3}{c}{LFI}&\multicolumn{6}{c}{HFI}\\ 
Characteristic\\ \hline
Frequency (GHz) &28.4&44.1&70.4&100&143&217&353&545&857\\
FWHM (arcmin)&32.3&27.0&13.2&9.7&7.2&4.9&4.9&4.7&4.2\\
Bandwidth (\%)&20&20&22&33&30&33&28&31&30\\
Sensitivity ($\mu K_{CMB}$ deg)&2.5&2.7&3.5&1.29&0.55&0.78&2.56&0.78$^*$&0.72$^*$\\ \hline
\end{tabular}
\caption{{\em Planck} instrument performance. Figures taken from the 2015 full mission data release \cite{p15_1}. $^*$ The sensitivity values for the 545 and 857 GHz channels, which are not sensitive to the CMB, are given in kJy sr$^{-1}$.}
\label{tab:inst}
\end{table}

The data processing required to take the observations made by the spacecraft and turn them into scientific results is complex and time consuming, and some elements of this processing for the most difficult tasks - large scale polarisation measurements - are still underway. Data and associated science papers have regularly been released (eg. \cite{p11_1}, \cite{pi_1}, \cite{p13_1}, \cite{p15_1}) with results either dated by year of release, or indicated as being early or intermediate results. The final results from the {\em Planck} Consortium are expected to be released later in 2017.

\section{The {\em Planck} View of the CMB}

\subsection{The Planck All Sky Maps}

\begin{figure}
\centering
\resizebox*{13cm}{!}{\includegraphics{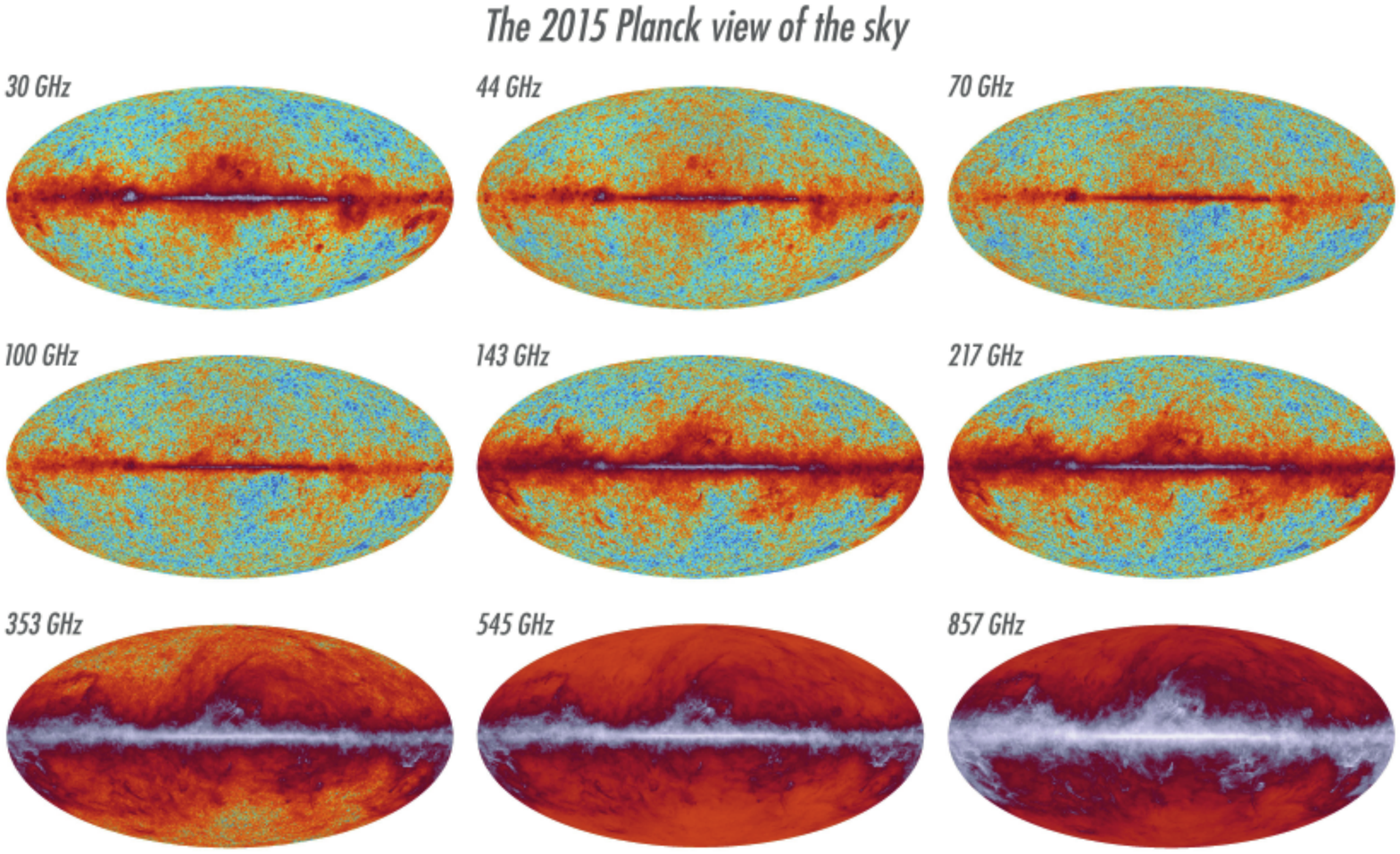}}\hspace{5pt}
\caption{The {\em Planck} maps of the whole sky in intensity at its nine different observing frequencies. These maps are presented in Galactic coordinates so the bright strip across the centre of each image is the Galactic plane. These images are projections of the entire sphere of the sky onto a flat plane using a Mollweide projection. Image courtesy of ESA.}
\label{fig:planck_sky}
\end{figure}

\begin{figure}
\centering
\resizebox*{14cm}{!}{\includegraphics{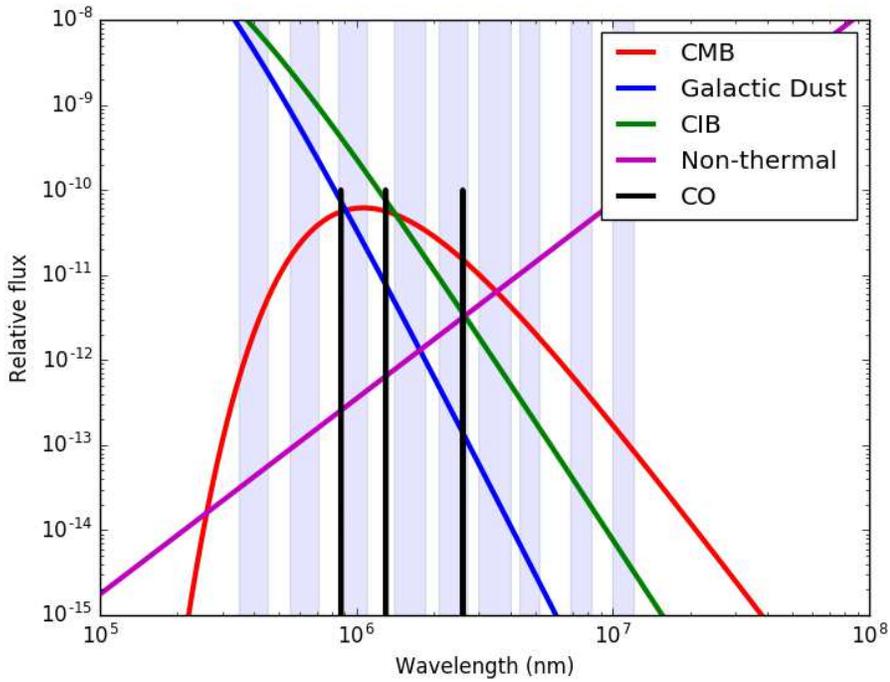}}\hspace{5pt}
\caption{The CMB spectral energy distribution (SED) compared to the various different foreground components. Red shows the CMB, blue shows the emission of dust in our own galaxy, green shows the cosmic infrared background (CIB; \cite{f98}) which dominates away from the galactic plane, magenta shows the shape of generic non-thermal emission which may come from the galactic plane or extragalactic sources. Black shows the position of the strong CO emissions lines. The strength of these different foreground components varies with position on the sky. The blue shaded regions show the wavelength coverage of the nine different {\em Planck} filters.}
\label{fig:planck_foregrounds}
\end{figure}

{\em Planck}'s basic data products are fully calibrated maps of the flux intensity received from the sky at its nine different observing frequencies, and the polarisation maps that are available for seven of these. This data is stored using a system called \verb!HEALPix! \cite{g05} which provides an equal area representation of the celestial sphere. These maps are available in a fully calibrated form  from the {\em Planck} Legacy Archive. The all-sky intensity maps are shown in Fig. \ref{fig:planck_sky}. These images, especially in the higher frequency HFI bands, are at significantly higher angular resolution and sensitivity than has been possible from any previous CMB space mission. However, these are maps of all the emission on the sky. They include emission not only from the CMB but from all other sources. The emission concentrated along the centre line of each of these maps, for example, comes from the plane of our own galaxy. Many different components contribute to these images, including non-thermal synchrotron and free-free emission, thermal dust emission, emission from carbon monoxide molecules and more (see Fig. \ref{fig:planck_foregrounds}). To be able to extract the emission of the CMB from these maps the data must go through a process known as component separation.

\subsection{Component Separation: Separating the CMB Wheat from the Astrophysical Chaff}

As can be seen from Fig. \ref{fig:planck_foregrounds}, the signal from the CMB is mixed with emission from a wide range of astrophysical components. These vary in strength from position to position, but, most importantly, they all have a very different dependence on wavelength. The most significant foreground contaminants are dust, which dominates at short wavelengths, and non-thermal emission which dominates at longer wavelengths.

Dust is a term that includes particles in the interstellar medium ranging in size from grains up to a micron across, to what are effectively large molecules. On Earth, these interstellar dust particles would be considered smoke or fumes since they are much smaller than the particles of household dust that we see every day. Interstellar dust is largely made up of carbon, silicon and oxygen. The dust emission seen by {\em Planck} is simply thermal emission from dust particles at temperatures of around 20 K in the interstellar medium of galaxies. This may come from within our own Galaxy, or be in the form of an integrated background, the Cosmic Infrared Background (CIB; \cite{f98}), that is the combination of dust emission from galaxies at all redshifts. This means that the CIB spectral energy distribution (SED) is somewhat flatter than for dust in our own Galaxy. In both cases, the contribution of dust falls with increasing wavelength in the Rayleigh-Jeans tail of thermal emission.

The non-thermal emission foreground dominates at longer wavelengths. This is emission arising from the acceleration of electrons inside the interstellar medium of our own and other galaxies. The two main processes behind this are: thermal Bremsstrahlung radiation, also known as free-free emission in astrophysics, whereby electrons and other charged particles scatter off each other; and synchrotron radiation, coming from electrons spiralling around magnetic fields. The SED of this long wavelength foreground can be characterised by a power law such that
\begin{equation}
S_{\nu} \propto \nu^{\beta}
\end{equation}
where $\beta$ ranges from $-2$ to $-4$ depending on the relative strength of the different emission processes.

As well as these sources of continuum radiation there are also strong lines from the carbon monoxide molecule that sit in, or close to, the passbands of three of the HFI channels.

The relative strength of these different components varies from place to place on the sky, but the broad parameters of their SEDs do not and, more importantly as can be seen from Fig \ref{fig:planck_foregrounds}, none of these components have an SED similar to that of the CMB. We can thus use observations in the nine different {\em Planck} bands to determine the contribution of each of these foregrounds at every position on the sky, and thus separate out the CMB from the foregrounds. This is  known as component separation.

\begin{figure}
\centering
\resizebox*{14cm}{!}{\includegraphics{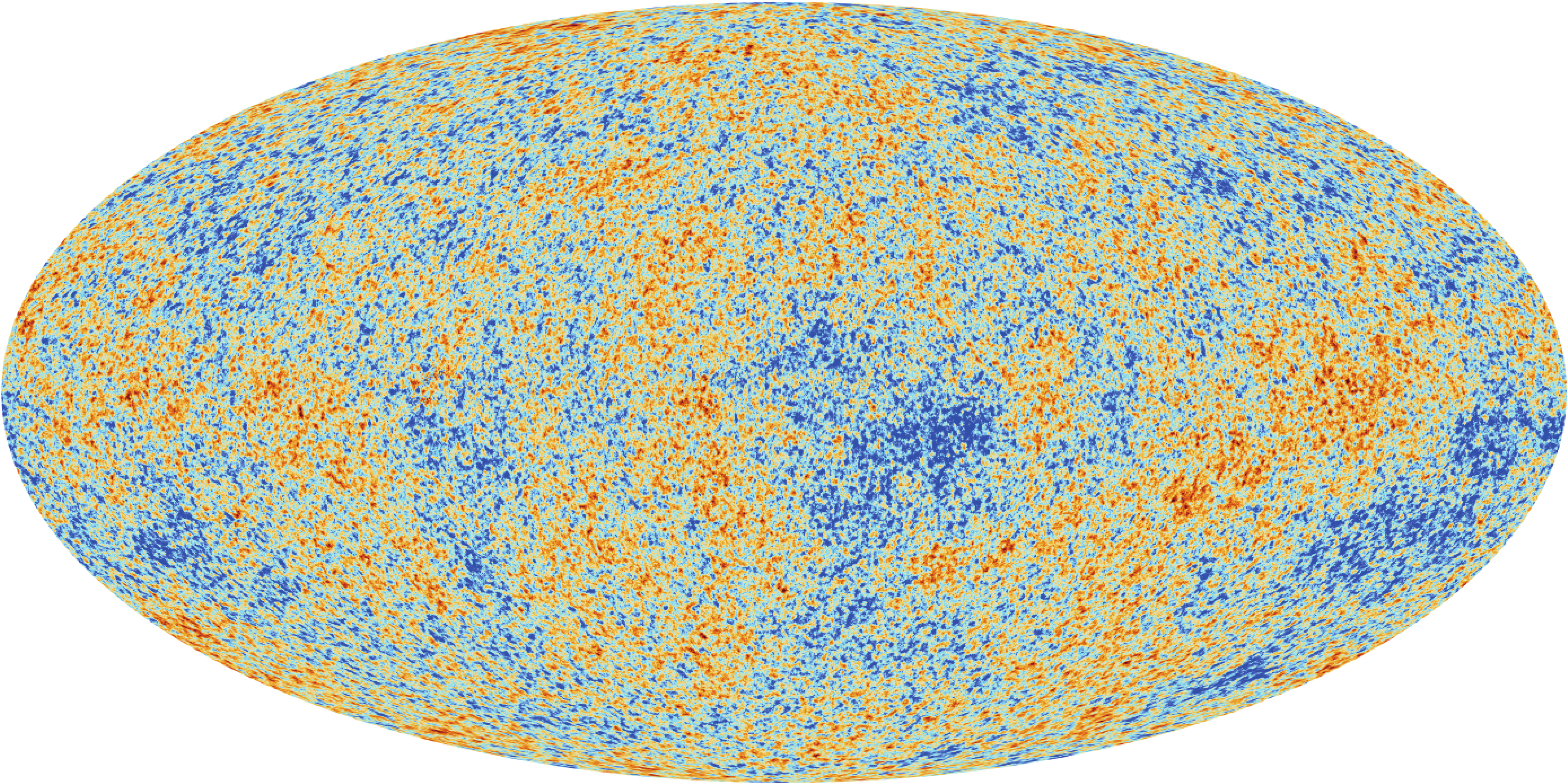}}\hspace{5pt}
\caption{The CMB as observed by {\em Planck}, revealing tiny temperature variations at the surface of last scattering. The colour scale on this image ranges from -300 (blue) to +300 (red) $\mu$K. Courtesy of ESA.}
\label{fig:cmb}
\end{figure}

Component separation is actually quite a complex problem which can be solved in a variety of different ways. The {\em Planck} project used four different approaches to component separation \cite{p13_12}, each of which used fundamentally different methods. Details of these can be found in \cite{p13_12}, and references therein. The results of these different approaches were then compared to make sure that a consistent separation of the foregrounds and measurement of the CMB was achieved. At the same time, individual sources, galactic and extragalactic, thermal or non-thermal, were also detected and removed from the maps.
The final result is a map of the CMB over the entire sky, shown in Fig. \ref{fig:cmb}, as well as maps and catalogs of the various different foregrounds.

\subsection{Power Spectrum Measurement}

\begin{figure}
\centering
\resizebox*{14cm}{!}{\includegraphics{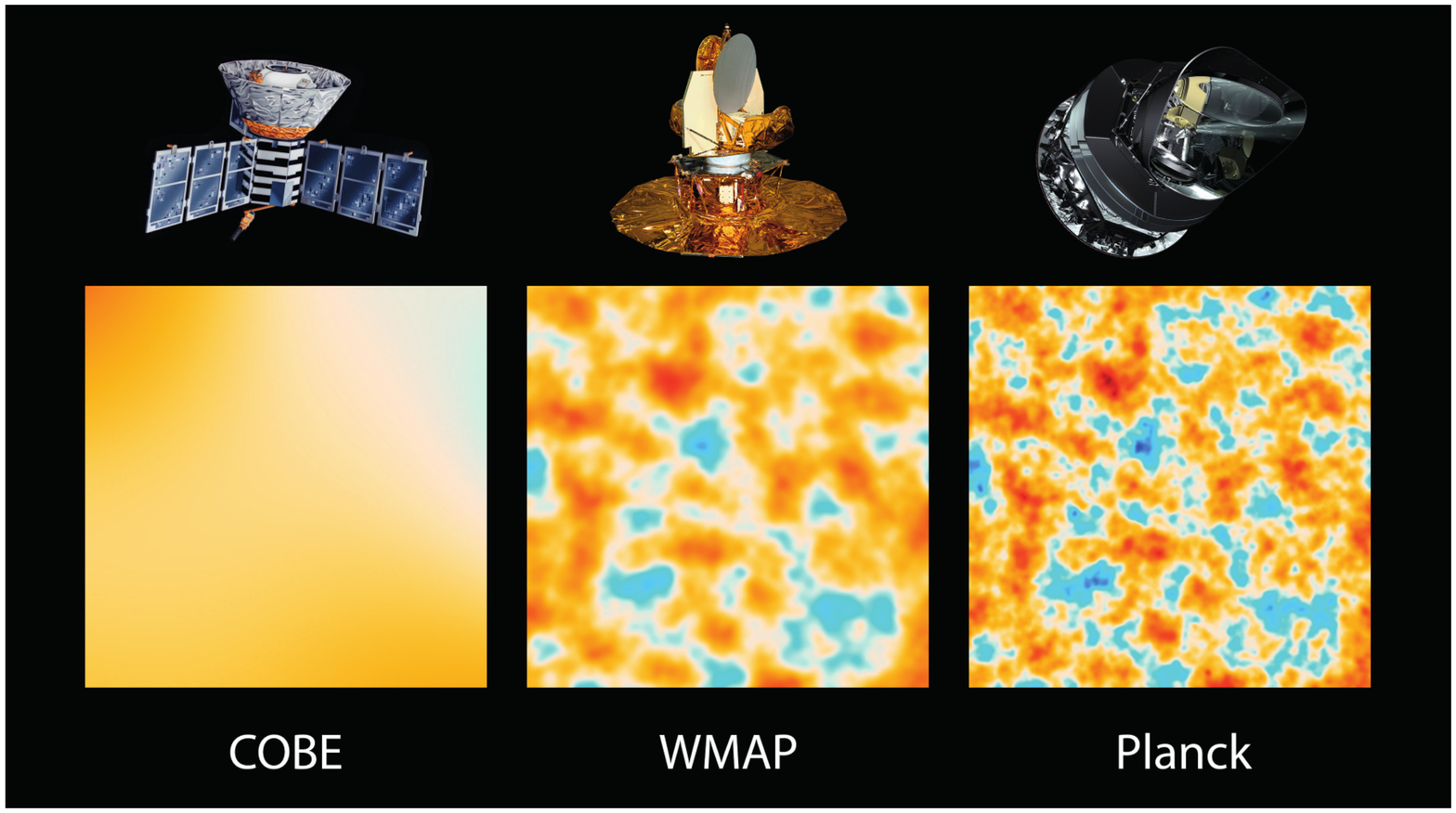}}\hspace{5pt}
\caption{A comparison of the angular resolution obtained by COBE, WMAP and {\em Planck} in mapping the CMB. Courtesy of NASA.}
\label{fig:res}
\end{figure}

To go from the {\em Planck} map of CMB temperature anisotropies seen in Fig. \ref{fig:cmb} to the cosmological parameters of the universe, we need to go from the individual details of the map, which is essentially a single realisation of many random processes, to a statistical measurement of the strength of temperature variations as a function of angular scale. This is a known as
the power spectrum of the CMB, and was shown in Fig. \ref{fig:planck_ps}. It is the equivalent of measuring the strength of different frequencies of sound using the square of a Fourier transform, but in two dimensions and on the surface of a sphere. While sine waves are the basis set for normal Fourier transforms, on the surface of a sphere we use spherical harmonics to form the appropriate orthonormal basis set. This set of functions are specified in a spherical polar coordinate system $(\theta, \phi)$ and are given by:
\begin{equation}
Y_l^m(\theta,\phi) = \left( -1 \right)^m \left [ \frac{(2l + 1)(l-m)!}{4 \pi (l+m)!}\right]^{1/2} P_l^m(cos ~\theta)\exp(im\phi)
\end{equation}
where $l$ and $m$ are integers with $l\geq 0$, $-l \leq m \leq l$ and $P_l^m$ is a Legendre polynomial specified by $l$ and $m$. The value of $l$ is termed the multipole, with higher values of $l$ corresponding to smaller angular scales, and thus higher multipoles. $l$ $\sim$100 corresponds to an angular scale of about 1 degree, near the peak of the anisotropy spectrum, while $l$ = 2 corresponds to a dipole. The high angular resolution of {\em Planck} allows multipoles up to 2500 to be measured (see Fig. \ref{fig:res}).

The power spectrum of the CMB can in principle be calculated by obtaining an autocorrelation function of the CMB map, in a manner analogous to that used to calculate the power spectrum of a sound wave, but this proves computationally difficult for the {\em Planck} data at small scales since the map involves of order $5 \times 10^7$ individual pixels. This is further complicated by the need to exclude pixels and regions affected by bright foreground sources, the larger scale effects of Galactic foregrounds, and the need to calculate uncertainties on each data point. Several statistical methods are used to calculate the power spectrum and its uncertainties (see \cite{p13_xv}, \cite{p15_xi}) which are checked against each other for consistency. Once this final observed CMB power spectrum is obtained it can be compared to the predictions of different cosmological models to reveal the nature of the universe.

\section{Planck Cosmology}

Comparison of cosmological models to the CMB anisotropy power spectrum from {\em Planck} allows many of the fundamental parameters of the universe to be derived \cite{p15_1}, \cite{p15_xi}, \cite{p15_13}. The precision of these parameters can be improved even further by combining the {\em Planck} data with results from other, complementary external data sets that look at other sources of cosmological information. These include baryon acoustic oscillations \cite{r15} and measurements of the distances and brightnesses of Type 1a supernovae \cite{c11}. The key results are described in Table \ref{tab:cos}, though the full {\em Planck} cosmological model includes a number of other, sometimes more technical parameters. See \cite{p15_1}, \cite{p15_13} and related {\em Planck} papers for more details.

\begin{table}
\begin{center}
\begin{tabular}{cc} \hline
Parameter&Value\\ \hline
\\
$H_0$ (kms$^{-1}$Mpc$^{-1}$)&67.74 $\pm$ 0.46\\
$\Omega_{\Lambda}$&0.6911 $\pm$ 0.0062\\
$\Omega_{b}$&0.04860 $\pm$ 0.00051\\
$\Omega_{c}$&0.2589 $\pm$ 0.0057\\
$\Omega_{m}$&0.3089 $\pm$ 0.0062\\
Age (Gyr)&13.799 $\pm$ 0.021\\
$n_s$&0.9667 $\pm$ 0.0040\\ \hline
\end{tabular}
\end{center}
\caption{The key cosmological parameters derived by combining the {\em Planck} observations with other existing cosmological data. $H_0$ is the Hubble constant, $\Omega_{\Lambda}$ is the fraction of the critical density of the universe that is produced by dark energy, $\Omega_{b}$ is the fraction produced by baryonic matter, $\Omega_{c}$ is the fraction produced by dark matter. $\Omega_{m}$ is the summation of these latter two terms and represents the fraction of closure density produced by matter. Since these terms all add up to 1 the geometry of the universe is flat. The Age of the universe is the time since the Big Bang, and $n$ is the slope of the initial perturbation spectrum. See \cite{p15_1}, \cite{p15_13} and references therein for more details.}
\label{tab:cos}
\end{table}

The first thing to note from Table \ref{tab:cos} is the unprecedented precision that {\em Planck} allows in the determination of cosmological parameters. For example, in the early 90s the Hubble constant, $H_0$, was not known to within a factor of two. {\em Planck} has now measured it to a precision of less than 1\%. The same is true for the age of the universe, which we now now to an accuracy of 20 million years. There are many aspects of the history of the Earth, for example the date of the emergence of life, that we know to far less precision.

The make up of the universe has also been made clear with these results. While cosmologists have long thought that the normal baryonic matter that makes up everything we can see is a largely insignificant constituent of the universe, the {\em Planck} results make that clear to very high precision. The matter we are made of makes up less than 5\% of the energy density of the universe, and just 15.7\% of the matter. The rest of the matter is made up of dark matter, that contributes gravitationally but which does not interact with electromagnetic radiation. While {\em Planck} is unable to determine the nature of this dark matter, it has clearly shown that it is by far the most dominant form of matter in the universe. It has also shown that matter, whether dark or baryonic, is not the dominant form of energy density in the universe. Nearly 70\% of this is produced by dark energy, the term in the Friedmann equation that Einstein is said to have thought his greatest mistake. Determining the physics behind dark matter and dark energy will be the job of 21st century science, whether at places like the Large Hadron Collider, through direct detection experiments in labs buried deep under mountains, or in space with next generation cosmological probes such as {\em Euclid} \cite{euclid}.

The other key parameter is the number $n_s$. This describes the size distribution of the initial density perturbations that led to the anisotropies in the CMB, and which are thought to arise from quantum noise during the epoch of inflation at very early times, when the universe was only a tiny fraction of a second old. The physics behind inflation is very poorly understood, and is likely to arise from fundamental physics processes that go well beyond the current standard model of particle physics. As such any information about these processes is very important. The value of $n_s$ determined by {\em Planck} corresponds to a slight deviation from scale free initial perturbations (which would have $n_s =1$, corresponding to an equal amount of power on all scales), which is consistent with the simplest `single field' models of inflation. There is also no evidence for curvature, also known as running, in this initial perturbation spectrum. This is also consistent with the simplest inflationary models.

While the analysis of the {\em Planck} cosmological results is considerably deeper and more detailed than can be presented here (see for example \cite{p15_13}), and while there is much more to be done in terms of matching these results to other data sets, the broad brush conclusion is that the universe appears to be much as we expected before {\em Planck}. This is a huge success. It shows that our understanding of the basic laws of physics, from the largest scales through General Relativity, to the microscopic scales through particle physics, is good enough to produce a highly accurate description of the universe from times as early as a fraction of a millisecond after the Big Bang right up to the present day. It also shows where gaps in our knowledge have to be filled, with the nature of dark energy and dark matter being central problems for the next decades. But the {\em Planck} results are also a bit of a disappointment - we have made no clear, unexpected discoveries, and appear, as one cosmologist put it, to be living in a `maximally boring' universe. But this is in fact a testament to the huge successes of observational cosmology. Since the discovery of the CMB we have gone from a completely mistaken view that the universe was steady state and unchanging, to a modern Big Bang cosmology that predicts the results of observations to an accuracy of 1\% or less.

Nevertheless, there are tensions between the {\em Planck} results and some other cosmological datasets. Currently the most significant of these  is a disagreement between the {\em Planck} value of the Hubble constant, $67.74 \pm 0.46$ and the value of $73.24 \pm 1.74$ from observations in the local universe largely obtained using the Hubble Space Telescope \cite{r16}. At face value this is a difference of 3$\sigma$, and is reaching the levels of significance where it starts to look real. The origins of any such disagreement could come from many directions. It might result from small inaccuracies in the details of modelling the variable stars and supernovae used in the local determination of $H_0$, or it might indicate that the strength of the dark energy term is not in fact constant but changes over time. Further observations are needed before this disagreement can be confirmed or understood. The nature and variability of dark energy, for example, will be the subject of ESA's next cosmology mission, {\em Euclid} \cite{euclid}.

There are also a few anomalies in the CMB maps themselves. The most significant of these have been named The Axis of Evil \cite{evil} and the Cold Spot \cite{cold}. The Axis of Evil is the finding that one half of the sky has a slightly higher CMB temperature than the other, and the split between the two is roughly aligned with the ecliptic plane of our own Solar System, while, the Cold Spot is a region of the CMB about five degrees across that is unusually cold. Both of these anomalies were originally discovered by {\em WMAP} but their statistical significance was questioned. They have both been independently confirmed by {\em Planck} so they are real, but their interpretation is unclear. The alignment of the Axis of Evil with the plane of our own Solar System suggests that it might be the result of some local foreground that has not been fully accounted for, but such a foreground has yet to be identified. Claimed explanations for the Cold Spot range from the possibility of a large scale void in the foreground galaxy distribution \cite{void}, to exotic physics such as cosmic textures \cite{texture} or colliding bubble universes \cite{bubble}. The nature of these and any other anomalies that may be lurking in the {\em Planck} data will be an active area of research for many years to come.

\section{{\em Planck} Foreground Science}

While the CMB was {\em Planck}'s main target, it also produced data on astrophysical objects across the entire universe which are having a huge impact on studies from the distant universe to our own galaxy. The papers that the {\em  Planck} Consortium has published in these non-cosmological areas are so numerous and varied that it is unfair to select any of the many dozens of results to highlight here. [A list of all {\rm Planck} papers can be found at \verb!https://www.cosmos.esa.int/web/planck/publications!]. Suffice it to say that {\em Planck} observations are having an impact over a wide range of astrophysics, from the nature and constituents of the interstellar medium of our own Galaxy, to galaxies and galaxy clusters at the highest redshifts.

{\em Planck} foreground observations have also already had an impact on subsequent cosmological observations. In early 2014 the team behind the BICEP2 experiment reported the detection of B-mode polarisation in the CMB. BICEP2 was an advanced CMB polarisation instrument observing a small region of the sky at a frequency of 150 GHz (a wavelength of 2 mm). Their detection of B-mode polarisation, if correct, would have been a major development, since it would be the first sign of the effect of gravitational waves on the CMB. This would both confirm that inflation had taken place in the very early universe, and be an indication of the physics behind it. The BICEP2 survey region was chosen to be a part of the sky that was low in continuum emission from dust. It was thought that this would also mean that foreground polarised dust emission would also be at a minimum. Since BICEP2 lacked observing channels at other frequencies it was not possible to use a component separation approach similar to that allowed by {\em Planck}'s nine separate observing bands. After the BICEP2 team announced these preliminary results at a press conference, the {\em Planck} team looked at their early maps of polarised foreground dust emission and found that there might be a problem \cite{pi_xxx}. Polarised dust emission covers the entire sky and is not necessarily at a minimum where the dust intensity is weakest. A joint {\em Planck}-BICEP2 team was established to look into this problem and concluded that the signal seen by BICEP2 was due to polarised dust and not to any effect in the CMB \cite{pb}. This incident demonstrates the very careful analysis needed in this field, as well as the value of the {\em Planck} data for future observations.

\section{The Next Steps}

The observational side of the {\em Planck} mission ended when the LFI instrument was finally switched off. Data analysis work continues, and the final data products and results on CMB polarisation are due later in 2017. The legacy of the mission will continue, and the all sky maps of the CMB and astrophysical foregrounds will be a resource for astronomers for many decades.

{\em Planck} has succeeded in its goal of extracting essentially all cosmological information available in the CMB intensity fluctuations. The next frontier for CMB astronomy, as mentioned above, is polarisation observations. {\em Planck} conducted polarisation observations, but the sensitivities obtained are unlikely to reach what is required to make a detection of the primordial B-mode fluctuations necessary to probe the physics of inflation. {\em Planck}'s final polarisation results will be released later in 2017. After this, the next step in CMB studies will be ground based experiments such as the Polarbear-2/Simons Array \cite{pb16} or balloon experiments like EBEX \cite{ebex17}. These both make use of large detector arrays, consisting of many thousands of detectors, to reach greater sensitivities than was possible with the small number of individual detectors on {\em Planck}.

Future space CMB missions have also been proposed for polarisation observations. The first of these likely to fly is LiteBIRD \cite{lb16}, a proposed Japanese mission to measure CMB polarisation on scales of tens of degrees. It would in many ways be to CMB polarisation studies what COBE was to intensity anisotropies. Larger, more complex missions have also been proposed, including the highly ambitious PRISM mission \cite{prism13} and the more {\em Planck}-sized CoRE mission \cite{core16}. Both of these were proposed to the European Space Agency (ESA) as part of the Cosmic Visions programme, with PRISM a candidate large mission and CoRE a candidate medium mission. Neither were selected for further study and development, but the potential remains for a future CMB mission that will do for polarisation and the inflationary epoch what {\em Planck} has done for CMB intensity fluctuations and the epoch of recombination. Whether the next developments are in space or on the ground, there is still much for us to learn from studies of the CMB.

\section{Conclusions}

{\em Planck} was ESA's mission to observe the CMB and to probe the cosmological parameters of the universe to unprecedented accuracy. To do this required substantial efforts by astrophysicists, instrument builders and spacecraft engineers over a period of nearly twenty years. The mission has been a great success, providing precision measurements of cosmological parameters to better than 1\%, as well as observations of astrophysical foregrounds that provide insights into nearly everything else in the universe. The data products produced by the {\em Planck} collaboration are freely available to anybody who wishes to use them. These, alongside the scientific results of {\em Planck}, which currently include at least 150 scientific papers, will be an astronomical legacy for many years.

\section*{Acknowledgements}

Based on observations obtained with Planck (\verb!http://www.esa.int/Planck!), an ESA science mission with instruments and contributions directly funded by ESA Member States, NASA, and Canada. The author wishes to thank A. Jaffe, J. Greenslade, and T-A. Cheng for useful discussions of this paper, as well as his many collaborators on the {\em Planck} project, without whom none of this would have been possible. The author would also like to thank Robert Wilson, co-discoverer of the CMB, for the use of his office during a visit to the Harvard-Smithsonian Centre for Astrophysics.


\begin{thebibliography}{99}


\bibitem{pw65}
Penzias~A.A., \& Wilson~R.W., A Measurement of Excess Antenna Temperature at 4080 Mc/s. ApJ. 1965; 142:419--421.

\bibitem{d65}
Dicke~R.H., Peebles,~P.J.E., Roll~P.G., Wilkinson~D.T. Cosmic Black-Body Radiation. ApJ. 1965;142: 414--419. 

\bibitem{h29}
Hubble~E., A Relation between Distance and Radial Velocity among Extra-Galactic Nebulae. PNAoS. 1929; 15: 168--173. 

\bibitem{t00}
Tytler~D., O'Meara~J.M., Suzuki~N, Lubin~D. Review of Big Bang Nucleosynthesis and Primordial Abundances. PhST. 2000; 85: 12-83.

\bibitem{p98}
Perlmutter~S. et al. Measurements of $\Omega$ and $\Lambda$ from 42 High-Redshift Supernovae. ApJ. 1999: 517; 565-586

\bibitem{infl}
Sriramkumar~L. An introduction to inflation and cosmological perturbation theory. Current Science. 2009: 97; 868-886.

\bibitem{l99}
Liddle~A.R. An introduction to cosmological inflation. In `High Energy Physics and Cosmology', 1999, eds. Masiero~A., Senjanovic~G., \& Smirnov~A., World Scientific Publishers, 1999: 260-295.

\bibitem{s68}
Silk~J. Cosmic Black-Body Radiation and Galaxy Formation. ApJ. 1968: 151; 459-471

\bibitem{m94}
Mather~J.C., et al. Measurement of the cosmic microwave background spectrum by the COBE FIRAS instrument. ApJ. 1994; 420: 439-444.

\bibitem{c76}
Corey~B.E., \& Wilkinson, D.T., A Measurement of the Cosmic Microwave Background Anisotropy at 19 GHz. 1976; BAAS, 8: 351.

\bibitem{h71}
Henry~P.S., Isotrropy of the 3K Background. Nature 1970;231:516-518.

\bibitem{s77}
Smoot~G. F, Gorenstein~M. V., Muller~R. A, Detection of anisotropy in the cosmic black body radiation. PRL. 1977; 39: 898-901.

\bibitem{s92}
Smoot~G. F, et al. Structure in the COBE differential microwave radiometer first-year maps. ApJL. 1992; 396; L1-L5.

\bibitem{m00}
Mauskopf~P. D., et al. Measurement of a Peak in the Cosmic Microwave Background Power Spectrum from the North American Test Flight of Boomerang. ApJL. 2000; 536: L59-L62.

\bibitem{pl06}
Planck Collaboration. The Scientific Programme of {\em Planck}. ESA publication ESA-SCI(2005)1. 2006: arXiv:astro-ph/0604069

\bibitem{h97}
Hu~W., \& White~M. A CMB Polarization Primer. New Astronomy. 2: 323-344.

\bibitem{k16}
Kamionkowski~M., \& Kovetz~E.D. The Quest for B Modes from Inflationary Gravitational Waves. ARAA. 2016; 54: 227-269.

\bibitem{p11_1}
Planck Collaboration. {\em Planck} early results I. The {\em Planck} mission. A\&A. 2011; 536: A1.

\bibitem{taub10}
Tauber~J.A., et al.. {\em Planck} pre-launch status: The {\em Planck} mission. A\&A. 2010; 520: 1-22.

\bibitem{m11}
Mennella~A., et al.. {\em Planck} early results III. First assessment of the Low Frequency Instrument in-flight performance. A\&A. 2011: 536: A3

\bibitem{p11_1}
Planck Collaboration. {\em Planck} early results II. The thermal performance of {\em Planck}. A\&A. 2011; 536: A2.

\bibitem{w04}
Werner~M.W., Gallagher~D.B., Irace~W.R. SIRTF - the Space Infrared Telescope Facility. AdSpR. 2004; 34: 600-609.

\bibitem{t95}
Thronson~H.A., Hawarden~T.G., Penny,~A.J., Vigroux~L., \& Sholomitskii~G. The Edison Infrared Space Observatory. SSR. 1995: 74: 139-144.

\bibitem{pi_1}
Planck Collaboration. Planck intermediate results. I. Further validation of new Planck clusters with XMM-Newton. A\&A. 2012: 543; A102

\bibitem{p13_1}
Planck Collaboration. Planck 2013 results. I. Overview of products and results. A\&A. 2014: 571; A1.

\bibitem{p15_1}
Planck Collaboration. Planck 2015 results. I. Overview of products and results. A\&A. 2016: 594; A1.

\bibitem{g05}
Gorski~K.M. et al.. HEALPix: A Framework for High-Resolution Discretization and Fast Analysis of Data Distributed on the Sphere. ApJ. 2005: 622; 759-771.

\bibitem{f98}
Fixsen~D.J. et al.. The Spectrum of the Extragalactic Far-Infrared Background from the COBE FIRAS Observations. ApJ. 1998: 508; 123-128.

\bibitem{p13_12}
Planck Collaboration. {\em Planck} 2013 results. XII. Diffuse component separation. A\&A. 2014: 571; A12.

\bibitem{p13_xv}
Planck Collaboration. {\em Planck} 2013 results. XV. CMB power spectra and likelihood. A\&A. 2014: 571; A15.

\bibitem{p15_xi}
Planck Collaboration. {\em Planck} 2015 results. XI. CMB power spectra, likelihoods, and robustness of parameters A\&A. 2016: 594; A11.

\bibitem{p15_13}
Planck Collaboration. {\em Planck} 2015 results. XIII. Cosmological parameters A\&A. 2016: 594; A13.

\bibitem{r15}
Ross~A.J., Samushia~L, Howlett~C, Percival~W.J., Burden~A., Manera~M. The clustering of the SDSS DR7 main Galaxy sample - I. A 4 per cent distance measure at z = 0.15. MNRAS. 2015: 449; 835-847.

\bibitem{c11}
Conley~A., Guy~J., Sullivan~M. Supernova Constraints and Systematic Uncertainties from the First Three Years of the Supernova Legacy Survey. ApJS. 2011: 192; 1-29.

\bibitem{r16}
Reiss~A.G., et al. A 2.4\% Determination of the Local Value of the Hubble Constant. ApJ. 2016: 826; 56--87.

\bibitem{euclid}
Laureijs~R., et al. Euclid: ESA's mission to map the geometry of the dark universe. Space Telescopes and Instrumentation 2012: Optical, Infrared, and Millimeter Wave. Proceedings of the SPIE. 2012: 8442; 84420-84428.

\bibitem{evil}
Land~K., \& Magueijo~J. Examination of Evidence for a Preferred Axis in the Cosmic Radiation Anisotropy. PhRvL. 2005: 95: 071301.

\bibitem{cold}
Cruz~M., Martínez-Gonzalez~E., Vielva~P, Cayon~L. Detection of a non-Gaussian spot in WMAP. MNRAS. 2005: 356: 29-40.

\bibitem{void}
Szapudi~I et al. Detection of a supervoid aligned with the cold spot of the cosmic microwave background. MNRAS. 2015: 450; 288-294.

\bibitem{texture}
Cruz~M et al. A Cosmic Microwave Background Feature Consistent with a Cosmic Texture. Science. 2007: 318; 1612-1614.

\bibitem{bubble}
Czech~B., et al. Polarizing bubble collisions. JCAP. 2010: 2010; 12.

\bibitem{pi_xxx}
Planck Collaboration. {\em Planck} Intermediate Results. XXX. The angular power spectrum of polarized dust emission at intermediate and high Galactic latitudes. A\&A. 2016: 586; A133.

\bibitem{pb}
Ade~P.A.R., et al. Joint Analysis of BICEP2/Keck Array and {\em Planck} Data. PRL. 2015: 114;101301.

\bibitem{pb16}
Suzuki~A., et al. The Polarbear-2 and the Simons Array Experiments. JoLTP. 2016: 185; 805-810.

\bibitem{ebex17}
The EBEX Collaboration. The EBEX Balloon-Borne Experiment - Gondola, Attitude Control, and Control Software. ApJSupp. 2017; in press.

\bibitem{lb16}
Matsumura~T., et al. LiteBIRD: Mission Overview and Focal Plane Layout. JoLTP. 2016: 184; 824-831.

\bibitem{prism13}
PRISM Collaboration. PRISM (Polarized Radiation Imaging and Spectroscopy Mission): A White Paper on the Ultimate Polarimetric Spectro-Imaging of the Microwave and Far-Infrared Sky. 2013: arXiv:1306.2259.

\bibitem{core16}
Di Valentino~E., et al. Exploring Cosmic Origins with CORE: Cosmological Parameters. 2016: arXiv:1612.00021.


\end{thebibliography}
\end{document}